\documentclass[a4paper]{article}
\usepackage[utf8]{inputenc}
\usepackage{authblk}
\usepackage{setspace}
\usepackage[margin=1.25in]{geometry}
\usepackage{graphicx}
\graphicspath{ {./figures/} }
\usepackage{subcaption}
\usepackage{amsmath}
\usepackage{xcolor}
\usepackage[english]{babel}
\usepackage[utf8]{inputenc}
\usepackage{multicol}

\usepackage[style=numeric,sorting=none]{biblatex}
\title{Geant4 Simulation  of an X-ray Beam Monitoring Detector based on plastic scintillators}

\author[1*]{C. H.~Zepeda~Fern\'andez}
\author[2]{E. Moreno~Barbosa}
\author[3]{J. M. Arredondo-Vel\'azquez}
\author[2]{L. F. Rebolledo-Herrera}

\affil[1]{SECIHTI--Facultad de Ciencias F\'isico Matem\'aticas,
Benem\'erita Universidad Aut\'onoma de Puebla,
Puebla 72570, Mexico}
\affil[2]{Facultad de Ciencias F\'isico Matem\'aticas,
Benem\'erita Universidad Aut\'onoma de Puebla,
Puebla 72570, Mexico}
\affil[3]{Cinvestav Unidad Tamaulipas,
Ciudad Madero, Tamaulipas, Mexico}
\affil[*]{Address correspondence to: hzepeda@fcfm.buap.mx}

\date{}

\begin{document}

\maketitle

\begin{abstract}

  Beam monitoring detectors play an important role in experimental control and radiological protection by enabling the real-time characterization of particle and photon beams. They provide information on beam intensity, position, and scattering, helping to ensure measurement accuracy, protect sensitive detectors, and quantify stray radiation in the surrounding environment.
In this work, a Geant4 simulation was developed to investigate the performance of an X-ray beam monitoring detector (BMD) composed of six hexagonal BC404 plastic scintillators arranged to provide high granularity, using a reflector with a realistic reflectivity of 93\%. Under the simulation conditions considered in this study, the detector exhibited an intrinsic time resolution of 100.8 $\pm$ 2.0~ps. The results indicated that only about 10\% of the incident photons interacted with the detector, suggesting that the proposed design could be used for beam characterization while minimally perturbing the primary beam. In addition, the X-ray multiplicity decreased with distance according to an inverse-square dependence. A backscatter study performed using a water phantom showed that approximately 6\% of the incident beam was detected as scattered radiation. These results suggest that the proposed detector may be a useful tool for X-ray beam characterization and for studies related to scatter radiation and radiological protection. Further experimental validation is required to confirm these findings.

\end{abstract}

\section{Introduction}\label{introduction}

When working with any type of beam, it is desirable to have a detector capable of monitoring it, as this enables the study of interaction and scattering processes. For example, in high-energy physics, beam monitors are used to identify and discriminate beam--beam minimum-bias and centrality events from background and beam--gas interactions. Plastic scintillator detectors have been widely employed for this purpose because they can generate minimum-bias trigger signals and provide information on beam activity~\cite{ALLEN2003549, ALICECollaboration_2008}. In addition, beam monitors help protect other detectors from damage caused by beam deviations~\cite{ATLAS, LHCb}. In high-radiation environments, these devices often require materials with excellent radiation tolerance, such as polycrystalline diamond grown by chemical vapor deposition (pCVD).
In low-energy applications such as X-rays, extreme radiation tolerance is generally not required. Instead, low-density materials are advantageous because they interact weakly with photons in this energy range. X-rays are commonly produced at synchrotron facilities, which typically require installations extending over hundreds of meters. More compact systems have also been developed, such as the Munich Compact Light Source (MuCLS) at the Technical University of Munich, a mini-synchrotron based on inverse Compton scattering (ICS) with dimensions of approximately 5~m $\times$ 3~m.  Such sources are used in biomedical imaging, including high-resolution computed tomography~\cite{Jacquet2016, Eggl2017, Paterno2020, Kulpe2020}. Scattered X-rays can degrade image quality and contribute to radiation exposure outside the primary beam; therefore, their characterization is important for image optimization, radiological protection, and shielding design~\cite{HARDING1999229,Seibert3}.  In materials science, X-rays are widely used to investigate crystal structures and internal properties~\cite{Huang2021, Huang2020, Melcher2024, Franck2021}. In all of these applications, beam monitors are useful for controlling and stabilizing the X-ray beam, whether generated by an ICS source~\cite{Gunther2023} or by a conventional X-ray tube.\\
Beam monitors are also used to evaluate beam intensity, linearity, spatial response, and detection efficiency, and to support the calibration of X-ray cameras and dosimetry systems. Detector technologies commonly employed for these purposes include solid-state materials such as CdTe, Si, and CZT, as well as scintillators such as LYSO and NaI~\cite{Habib2015, Hsieh2016, Crespo2016, Iniewski2011}. When beam monitoring is performed during an experiment, the detector should interact as little as possible with the incident X-rays in order to minimize perturbations to the primary beam. For example, monitoring of ICS sources requires highly stable and precise detectors~\cite{Gunther2019}.\\
Plastic scintillators offer several advantages in applications where minimal beam perturbation and fast temporal response are desirable. Owing to their low effective atomic number, they interact weakly with X-rays, resulting in lower detection efficiency and poorer energy resolution than inorganic scintillators~\cite{D1RA01878G, Sharma2023_JPETEfficiency, Siwal_2025}. At the same time, this property allows them to behave as quasi-transparent detectors, which makes them attractive for beam monitoring and dosimetric applications~\cite{vaneijk2002medical, saintgobain_plastic}. In addition, plastic scintillators exhibit short decay times, good mechanical robustness, relatively low cost, and considerable flexibility in terms of geometry and scalability~\cite{Sharma2023_JPETEfficiency, ZHANG2024165247, vaneijk2002, PDG2020, LI2005449, yanagida2018}.\\
Time resolution (TR) is one of the most important characteristics of a radiation detector and represents the minimum time interval required to distinguish two temporally separated events ($\sigma_{\mathrm{TR}}$). In experimental measurements, the overall timing performance depends not only on the detector itself, but also on contributions from the photodetector, front-end electronics, and signal processing. The intrinsic time resolution (ITR) corresponds to the fundamental timing limit imposed solely by the scintillator detector, excluding external contributions such as photodetector transit time spread and electronic jitter. In scintillation detectors, the ITR is governed by the stochastic nature of scintillation photon production, optical photon transport, reflections, absorption, detector geometry, and the effective photosensitive area. For a telescope-like setup composed of two detectors, A and B, where the ITR of detector B ($\sigma_{B}$) and the electronic contribution ($\sigma_{\mathrm{ele}}$) are known, and the ITR of detector A ($\sigma_{A}$) is unknown, the overall time resolution can be expressed as $\sigma_{\mathrm{TR}}^{2} = \sigma_{A}^{2} + \sigma_{B}^{2} + \sigma_{\mathrm{ele}}^{2}$.\\
In the present work, the ITR was estimated from the temporal distribution of detected optical photons. The most probable values obtained from the photon arrival time distributions were used to construct a global time distribution, which was fitted with a Gaussian function. The standard deviation of the fit was taken as the ITR because it quantifies the statistical spread of the detection times and, therefore, the minimum timing uncertainty associated exclusively with the scintillator and its optical properties (see Section~\ref{ITR}).\\
 In this work, an X-ray beam monitoring detector (BMD) based on hexagonal BC404 plastic scintillator cells is proposed. Owing to the low effective atomic number of this material, X-rays interact only weakly with the detector, which is advantageous for beam monitoring applications where minimal perturbation of the primary beam is desired. The hexagonal geometry allows the active area to be expanded by adding additional cells, thereby providing increased granularity. In addition, the compact dimensions of the detector make it suitable for portable applications. The remainder of this paper is organized as follows. Section~\ref{BMDConf} describes the configuration used as the basis for the BMD design. Section~\ref{simulation} presents the optical simulation of the BC404 plastic scintillator and the X-ray beam model employed in this study. Section~\ref{ITR} describes the characterization of the hexagonal cell in terms of its intrinsic time resolution (ITR) and the evaluation of its X-ray detection efficiency. Section~\ref{monitoring} presents the X-ray multiplicity predicted by the BMD as a function of distance. An example of a backscattering analysis is presented in Section~\ref{backscatter}. Section~\ref{uncertainty} discusses the statistical and systematic uncertainties, as well as the principal limitations of the simulation model. Finally, Section~\ref{conclusions} summarizes the main findings and conclusions of this work.

  \section{The BMD configuration}~\label{BMDConf}
\begin{figure}
 \includegraphics[width=\linewidth]{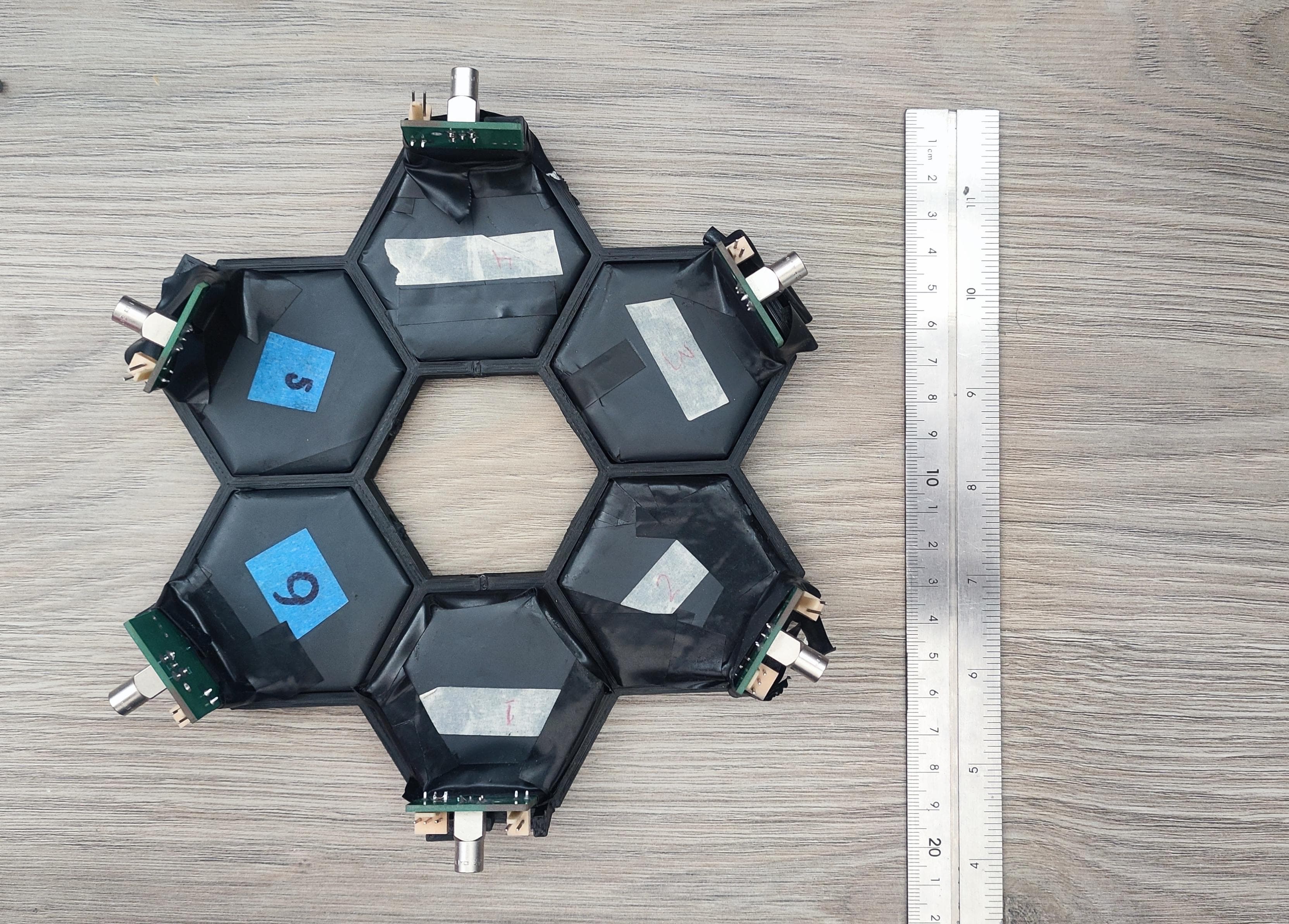}
 \caption{Frontal view of the detector showing the six reflectively wrapped and enclosed hexagonal scintillators, the surrounding plastic support, and the overall size indicated by a ruler in cm.}
 \label{BMD}
\end{figure}

The BMD geometry consists of an array of six concentric plastic scintillator hexagonal BC404 cells forming the first ring of the detector. Owing to the modular nature of the hexagonal design, additional rings can be incorporated to increase the active area and granularity as required. Each scintillator cell has a height of 5~cm and a thickness of 2~cm. The BMD was developed as part of a doctoral dissertation~\cite{Marquez2024}. In the experimental prototype, a SensL MicroFC-60035-SMT silicon photomultiplier (SiPM) with an active area of $6 \times 6$~mm$^2$ was used as the photosensor~\cite{SensL}. A photograph of the detector is shown in Figure~\ref{BMD}. The central opening of the detector is designed to surround an X-ray beam or beam pipe, allowing the system to monitor beam intensity and to study phenomena such as centrality and beam--gas interactions. A larger version of this concept, known as the Beam--Beam Monitoring Detector, incorporates six additional rings to extend the active area and provide higher granularity~\cite{bebe}. The number of rings can therefore be adapted according to the requirements of a particular beam-monitoring application.

\section{Simulation methodology}~\label{simulation}
\subsection{BC404 simulation material}
\begin{figure}
 \includegraphics[width=\linewidth]{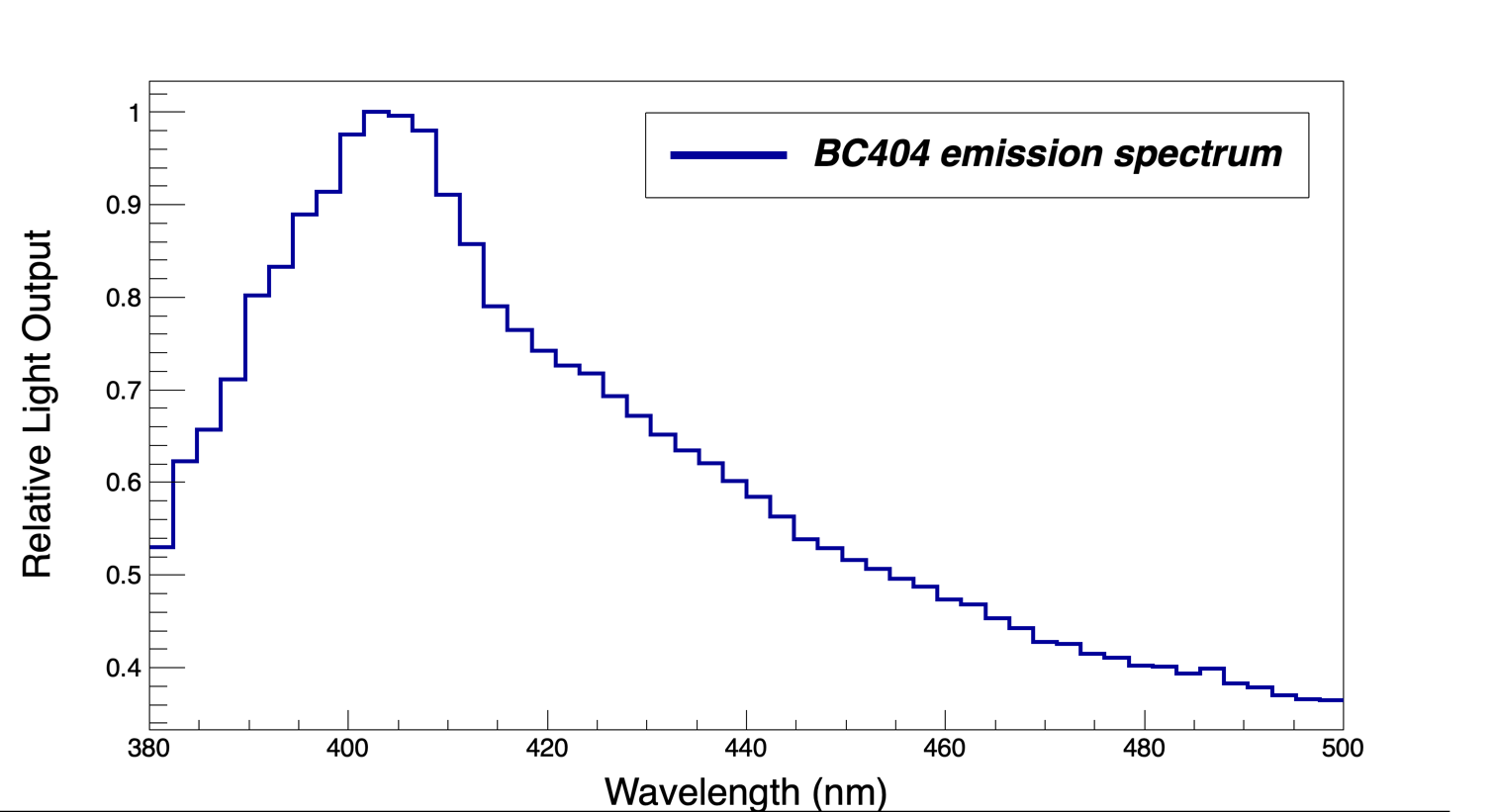}
 \caption{BC404 plastic scintillator emission spectrum obtained from simulation.}
 \label{spectrumbc404}
\end{figure}

All simulations in this  study were carried out using the GEANT-4 toolkit~\cite{Geant4}, version \textit{geant4-v11.1.3}, which provides a comprehesive framework for modeling particle-matter interactions. The hexagon was simulated with the dimension previously mentioned. The base material of BC404, polyvinyltoluene, was simulated by \textit{G4\_PLASTIC\_SC\_VINYLTOLUENE}, chosen from the GEANT-4 materials list. In Table~\ref{characteristicBC404} the optical simulated properties are shown, according to the BC404 data sheet~\cite{BC404}, while the light output property was taken from a Doctoral Dissertation~\cite{lightouput}. With this information the spectrum emission simulated is shown in Figure~\ref{spectrumbc404}. The GEANT-4 environment models most of the optical photon physics in the detector, such as ionization process, Compton, photoelectric, Cherenkov and photon propagation~\cite{Korpachev_2017}; all of which were utilized in this work. The effective area of the SiPM coupled to each hexagonal cell was modeled as a  $6\times6$~mm$^2$ square, which is referred to as the \textit{scorer} and was simulated using \textit{G4\_GLASS\_PLATE}. Both, were considered in an air enviroment, simulated using \textit{G4\_Air}. The surrounding support shown in Figure~\ref{BMD} was also simulated using polylactic acid (PLA) with a density of 1.24 g/cm$^3$ and a chemical composition of C\textsubscript{3}H\textsubscript{4}O\textsubscript{2}~\cite{Ranakoti2022_PLA_review}. Due to the supporting structure, the separation distance between the hexagons is 4~mm.

\begin{table}
 \centering
 \caption{Optical properties for BC404 simulation}
 \begin{tabular}{c | c | c}
Light output & Refraction index & Light atenuation\\
  \hline
10,880 &  1.58 & 1.4  m\\
photons/MeV & &\\
 \end{tabular}\label{characteristicBC404}
\end{table}

\subsection{Boundary conditions}\label{boundaryconditions}
For this study, two boundary conditions were considered, as explained below:
\begin{itemize}
\item Scintillator--Air surface: This optical boundary was modeled as a polished dielectric--metal interface with a reflectivity of 93\%, representative of commonly used reflective wrapping materials such as Mylar. This assumption was adopted to provide a more realistic description of the optical boundary conditions and to reduce the escape of optical photons from the plastic scintillator.
\item Scintillator--scorer surface: This optical boundary was modeled as a dielectric--metal interface with 100\% absorption. Under this condition, optical photons reaching the scorer surface were terminated and recorded only once, preventing any subsequent reflection or multiple counting.
\end{itemize}
Therefore, the width of the scorer is irrelevant, since no optical photons travel through its volume, and no additional boundary conditions are required for the scorer or the environment.
\subsection{X-ray spectrum simulation}\label{tungsten}
\begin{figure}
 \includegraphics[width=\linewidth]{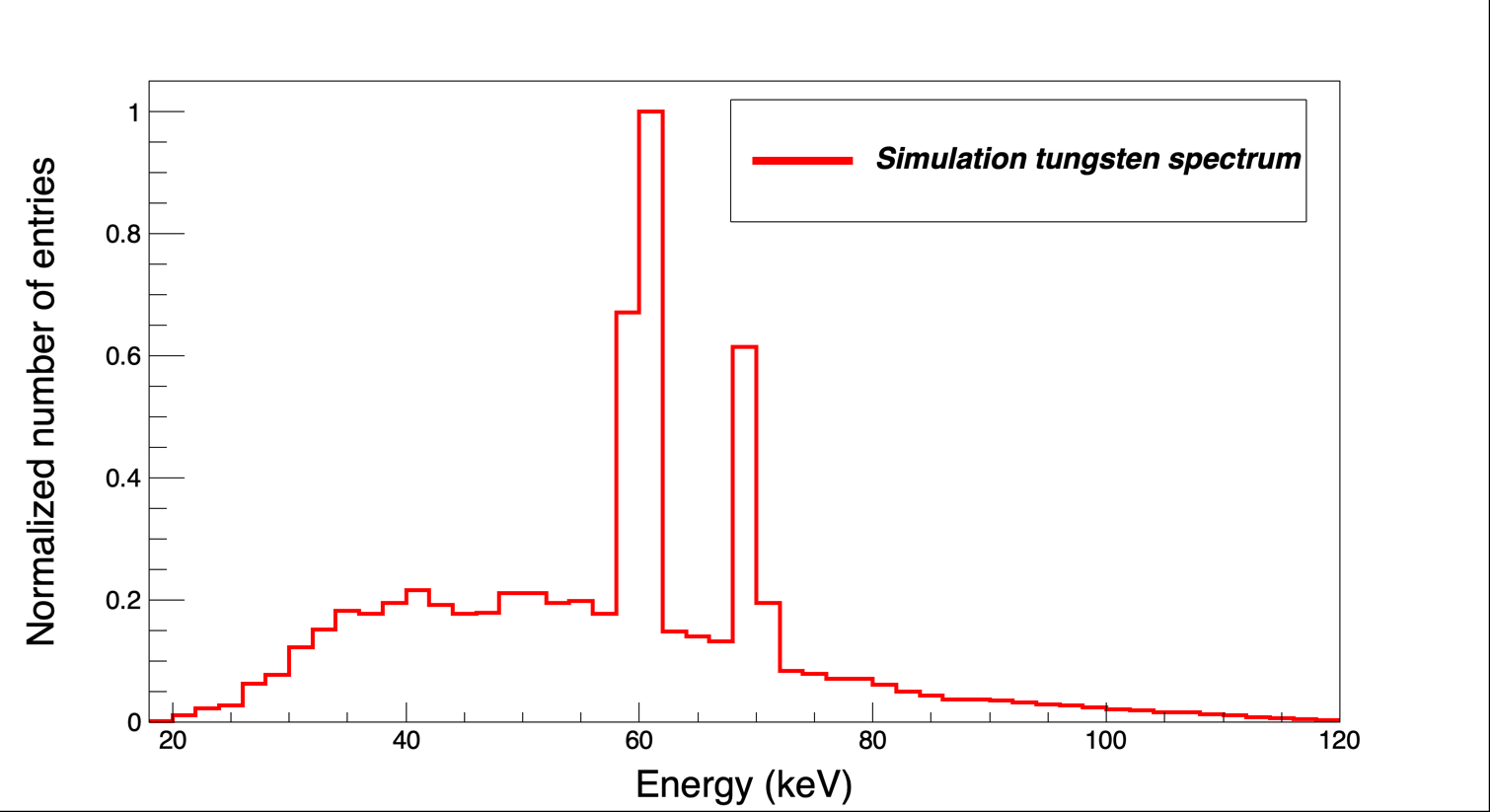}
 \caption{Simulated tungsten X-ray emission spectrum.}
 \label{tungstenspectrum}
\end{figure}

In X-ray generators, tungsten is the most commonly used target material. It is well known that only about 1\% of the X-ray photons are produced, while the remaining 99\% of the electron energy is converted into heat. Performing a simulation of creating an X-ray from an electron beam can be very time-consuming~\cite{Chatzisavvas2022XrayTubeMC, SALVAT20061201, Alijani2021_XraySourcesMC}. However, there is specialized software that allows this process or can simulate medical images~\cite{Salvat2019PENELOPE2018, Sloth2024McXtraceGPU, Kochebina2024OpenGATE}. Since the objective of this work is not the detailed simulation of X-ray production, a predefined X-ray spectrum was used in the simulations. The energy distribution was taken from ~\cite{tungsten}, where the bremsstrahlung and characteristic radiation produced by electrons incident on a tungsten target operated at 120 kV were calculated. The spectrum is shown in Figure~\ref{tungstenspectrum}, consisted of $3\times10^8$ photons. The photon emission positions were sampled from a uniform spatial distribution within a 2~mm $\times$ 2~mm window. The initial photon directions were generated assuming a uniform angular distribution constrained by the geometrical aperture defined by the source-detector configuration. This simplified source model provides a first-order approximation of the incident beam and allows the detector response to be evaluated without explicitly simulating the X-ray tube. However, the assumptions of a uniform spatial and angular distribution do not fully reproduce the characteristics of a realistic X-ray beam, and variations in beam divergence and source geometry may affect the quantitative detector response.

  \section{Intrinsic time resolution measurements for BMD}\label{ITR}
\begin{figure}
      \includegraphics[width=\linewidth]{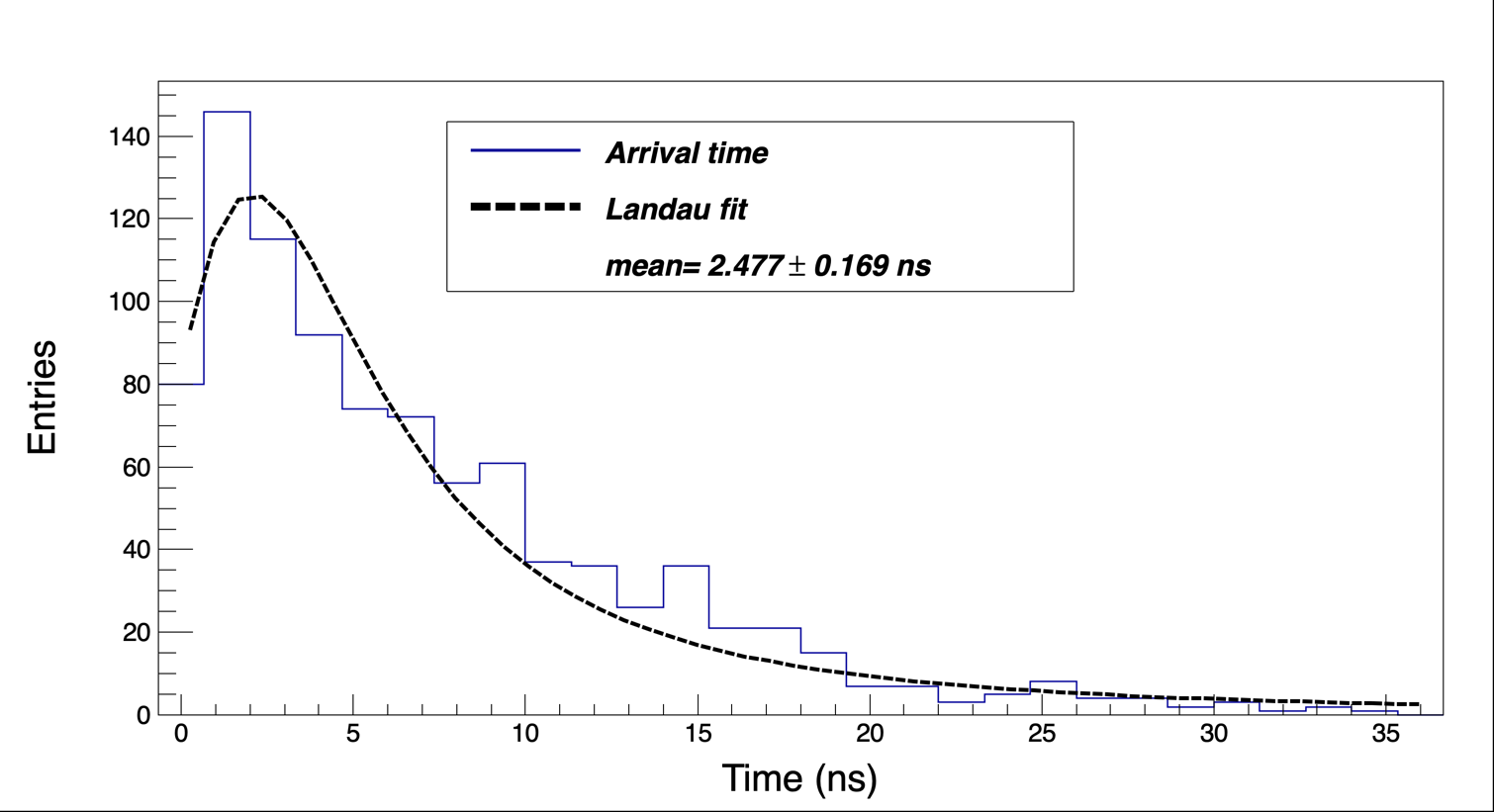}
      \caption{Arrival-time distribution of optical photons fitted with a Landau function.}
      \label{landau}
    \end{figure}

\begin{figure}
      \includegraphics[width=\linewidth]{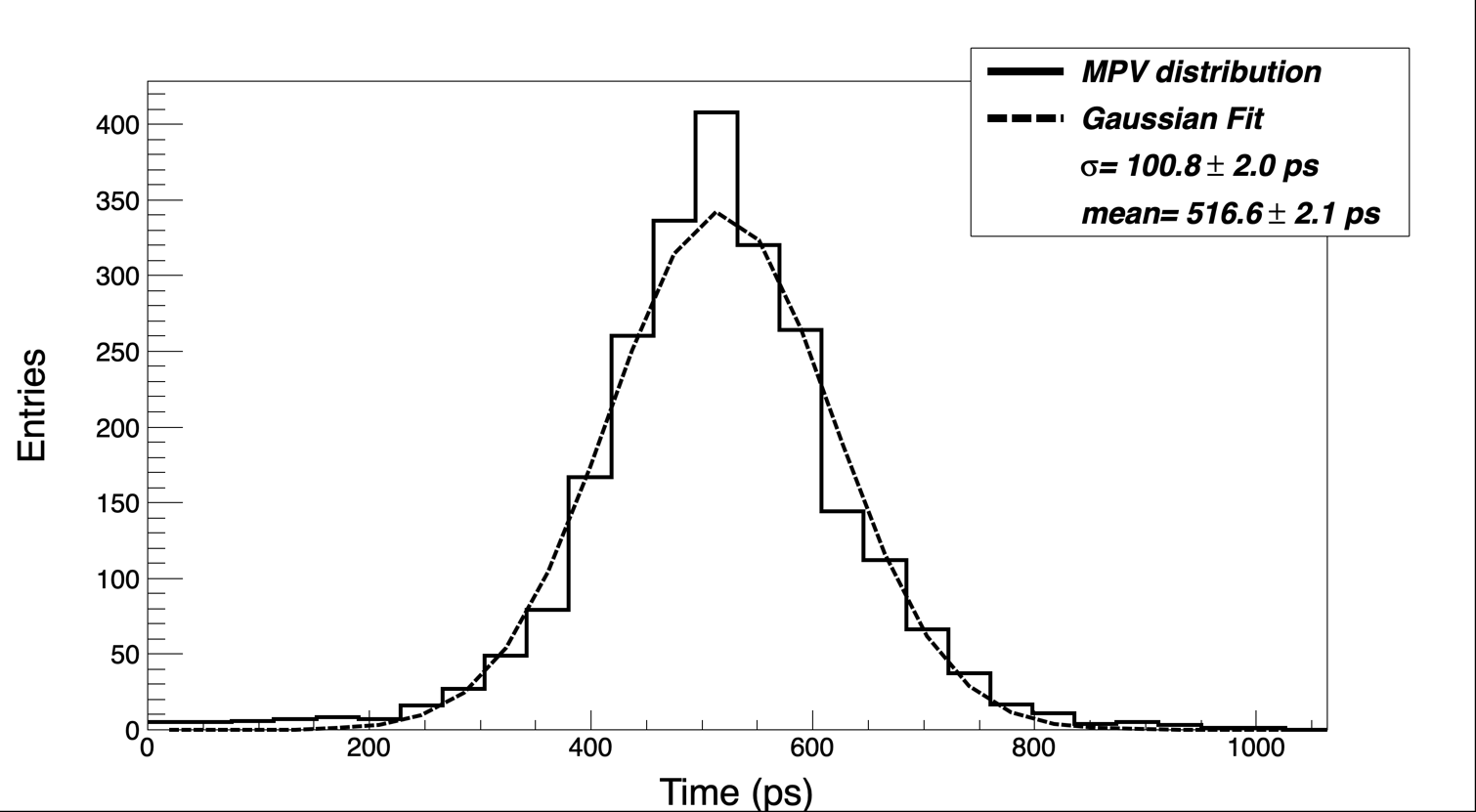}
      \caption{$\mu$ distribution fitted with a Gaussian function to determine the Intrinsic Time Resolution (ITR).}
      \label{ITRDistribution}
    \end{figure}

  A study of the ITR for a hexagonal cell at high energy was previously carried out in a beam test using 5~GeV pions, where a value of 45~ps was obtained by considering the section corresponding to the most probable charge value. A corresponding simulation was also performed using Geant4, and the resulting ITR was found to be consistent with the experimental measurement~\cite{bebe}. The same technique was adopted in the present work and is described below.\\
  The X-ray beam follows the profile shown in Figure~\ref{tungstenspectrum} and interactes at the geometric center of the hexagonal cell. For each interaction of an X-ray photon, on average  $150\pm13$ optical photons were detected by the scorer.\\
In a real photosensor, not every incident photon generates a detectable signal because the photon detection efficiency is lower than 100\%. Therefore, the trigger time for each event is defined as the most probable value (MPV, denoted by $\mu$) of the photon arrival time distribution, being the number of photons required to activate the photosensor.  An example of the arrival time distribution is shown in Figure~\ref{landau}. As a result, highlighting the possible BMD’s performance as a beam monitor, it was found that only about 10\% of the X-ray photons interact with the scintillator. Therefore, the BMD perturbs the beam by less than 10\%, ensuring minimal disturbance to its original profile. Experimental measurements are required to compare with this data.\\
The MPV obtained from the photon arrival time distributions for each X-ray photon interaction were used to construct a global time distribution, which is shown in Figure~\ref{ITRDistribution}. A Gaussian fit was performed yielded an ITR of  100.8 $\pm$2.1~ps. The standard deviation ($\sigma$) of the fit was taken as the ITR since it represents the statistical spread of the detection times and, consequently, the minimum timing uncertainty associated solely with the detector.\\
For comparison, an additional idealized simulation was performed assuming a reflector with 100\% reflectivity. Under this assumption,  710$\pm$31 optical photons reached the scorer per event, compared with  150$\pm$13 photons per event when a more realistic reflectivity of 93\% was used. Despite the larger number of detected photons, ITR increased from 100.8$\pm$2.0~ps to 199.3$\pm$4.1~ps. This behavior is attributed to the greater number of optical photons undergoing multiple reflections and eventually reaching the scorer, which broadens the arrival-time distribution and increases its mean value from approximately 516.6$\pm$2.1~ps to 2,745.7$\pm$5.6~ps. In contrast, when a reflectivity of 93\% is used, a fraction of the optical photons is lost at each reflection, either through absorption in the reflective material or by escaping the scintillator. This preferentially suppresses late-arriving photons, reducing the temporal spread and improving the intrinsic time resolution (ITR).  This comparison highlights the strong influence of reflector properties on the timing performance of the detector and suggests that maximizing light collection does not necessarily lead to optimal timing performance. A previous study on the modeling of scintillation detectors has shown that the survival probability of an optical photon after undergoing multiple reflections can be expressed as $P = R^{N}$, where $R$ is the reflectivity of the wrapping material and $N$ is the number of reflections experienced by the photon~\cite{Roncali2017}. This simple relationship illustrates how small reductions in reflectivity can lead to significant photon losses when long optical paths are involved. For the two reflectivity values considered in this work (93\% and 100\%), selected survival probability values are presented in Table~\ref{prob} to illustrate this trend. As expected, the 93\% reflectivity case yields substantially lower survival probabilities for photons undergoing multiple reflections. This behavior is consistent with the smaller number of optical photons recorded in the scorer and supports the observed reduction in the ITR, since late-arriving photons are preferentially lost rather than reaching the scorer.  Further studies are required to confirm this behavior and to determine the optimal reflector characteristics for this application.\\
Having characterized the timing performance of each scintillator cell, the X-ray beam monitoring results are presented in the following section.

\begin{table}
\centering
\caption{Optical photon survival probability, $P = R^{N}$, for reflectivities of 93\% and 100\%, where $R$ is the reflectivity and $N$ is the number of reflections experienced by the photon.}
\begin{tabular}{c | c | c}
$N$  & $P$ ($R = 93\%$) & $P$ ($R = 100\%$) \\
\hline
1  & 0.9300 & 1.0000 \\
2  & 0.8649 & 1.0000 \\
3  & 0.8044 & 1.0000 \\
4  & 0.7481 & 1.0000 \\
5  & 0.6957 & 1.0000 \\
10 & 0.4840 & 1.0000 \\
20 & 0.2342 & 1.0000 \\
\hline
\end{tabular}\label{prob}
\end{table}


  \section{Monitoring simulation tungsten X-ray profile}\label{monitoring}
  
\begin{figure}
        \includegraphics[width=\linewidth]{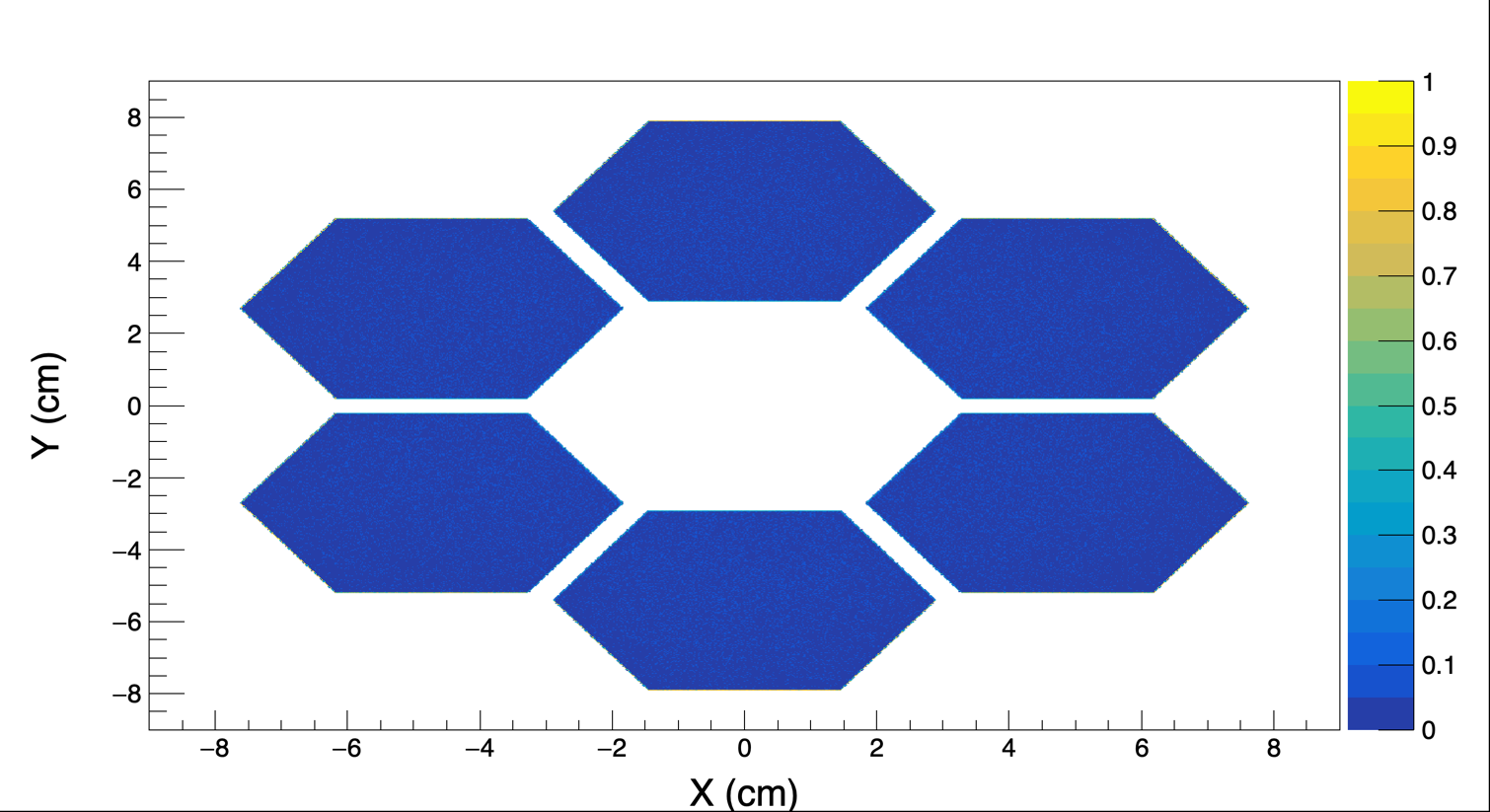}
        \includegraphics[width=\linewidth]{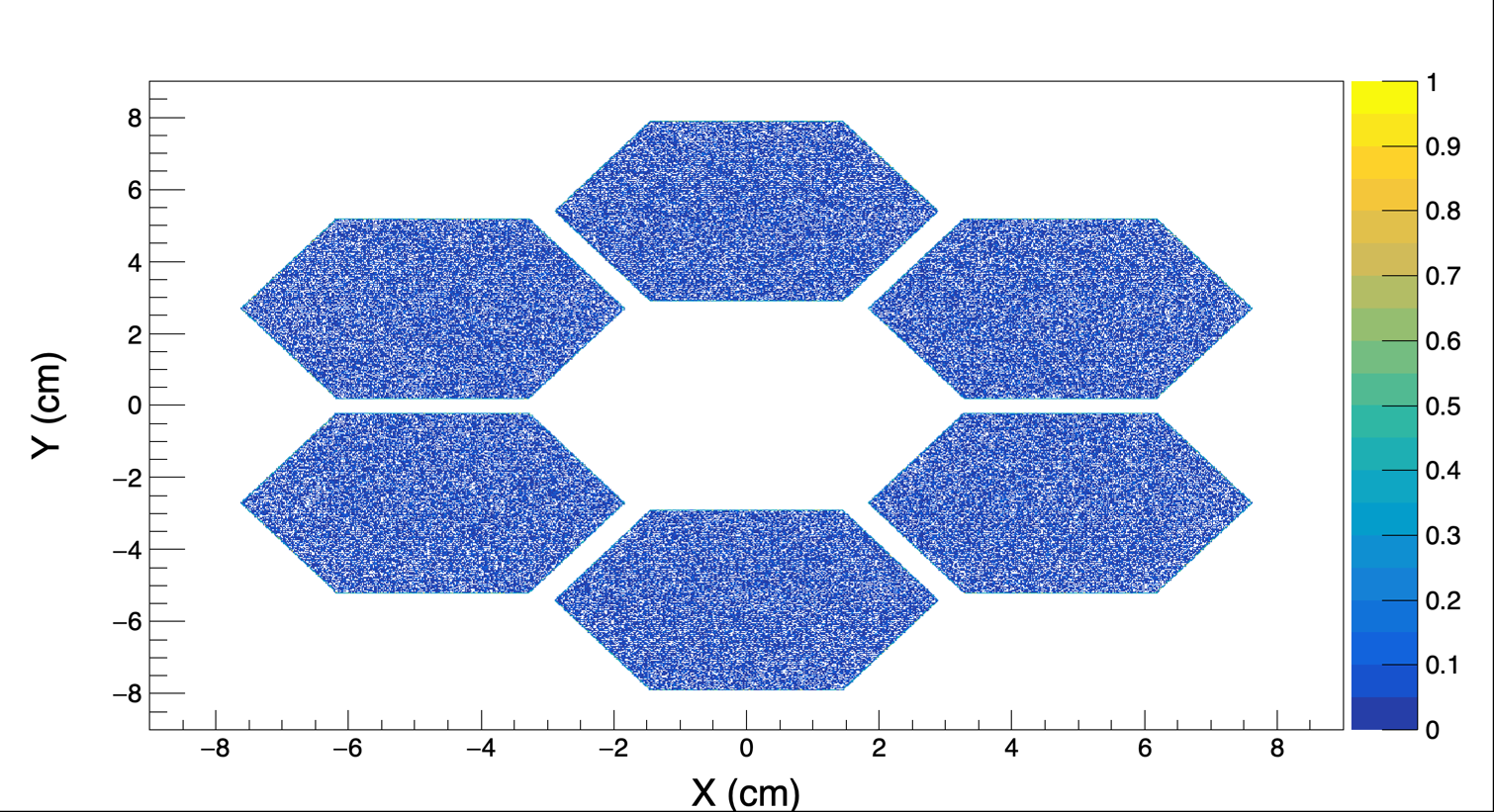}
        \caption{Qualitative comparison of the detected X-ray multiplicity for distances of 50~cm (up) and 200~cm(bottom).}
        \label{50cm}
      \end{figure}

\begin{figure}
        \includegraphics[width=\linewidth]{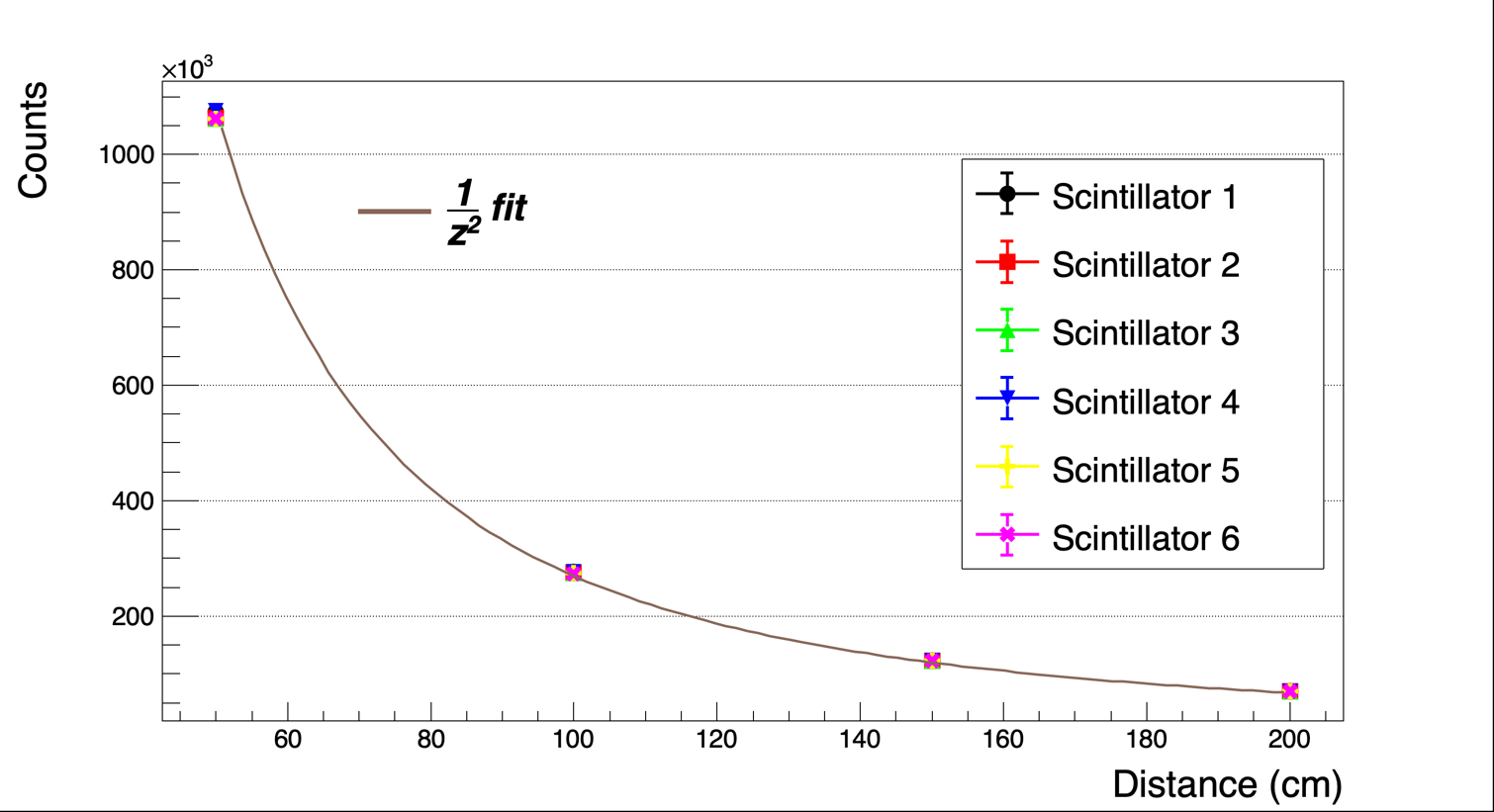}
        \caption{Distance dependence of the X-ray multiplicity, which exhibits a decay behavior proportional to $\sim$1/z$^2$.}
        \label{decay}
      \end{figure}

The hexagonal cells that constitute the BMD were numbered clockwise, starting with the upper cell as cell 1. In this simulation, the same X-ray profile described in Subsection~\ref{tungsten} was used. The BMD was placed at four positions along the \textit{z}-axis, located 50~cm, 100~cm, 150~cm, and 200~cm from the X-ray source, with the detector oriented in the \textit{xy}-plane.\\
First, the X-ray multiplicity was determined for each cell. Figure~\ref{50cm} presents a qualitative comparison of the multiplicity distribution in the six BMD cells for two distances: 50~cm and 200~cm. At each position, the distribution is approximately uniform across the active detector area, corresponding to the hexagonal cells. As expected, the multiplicity decreases with increasing distance from the source. This dependence is shown in Figure~\ref{decay}, where the decrease follows the inverse-square law.


\begin{figure}
        \includegraphics[width=\linewidth]{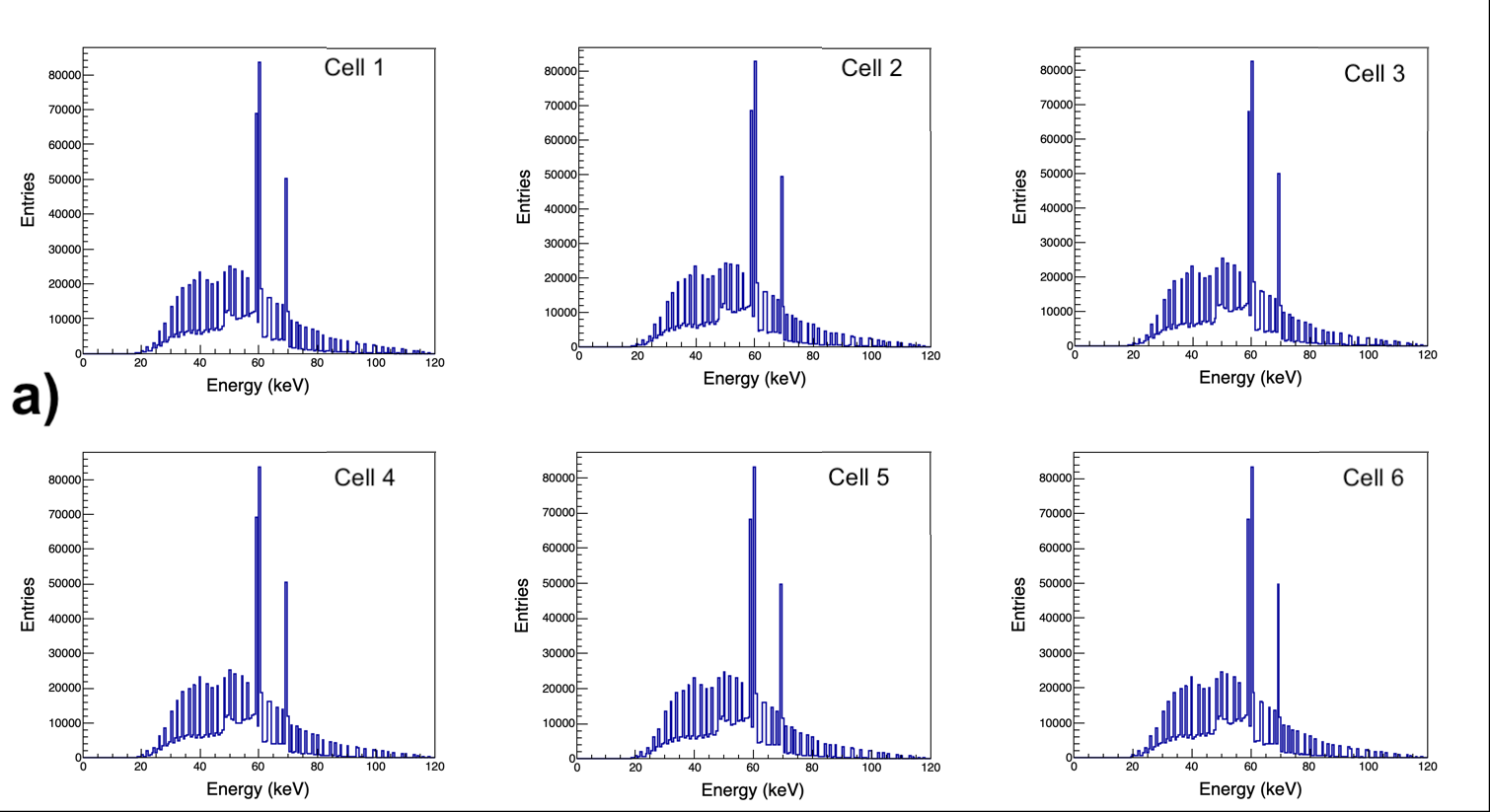}
        \includegraphics[width=\linewidth]{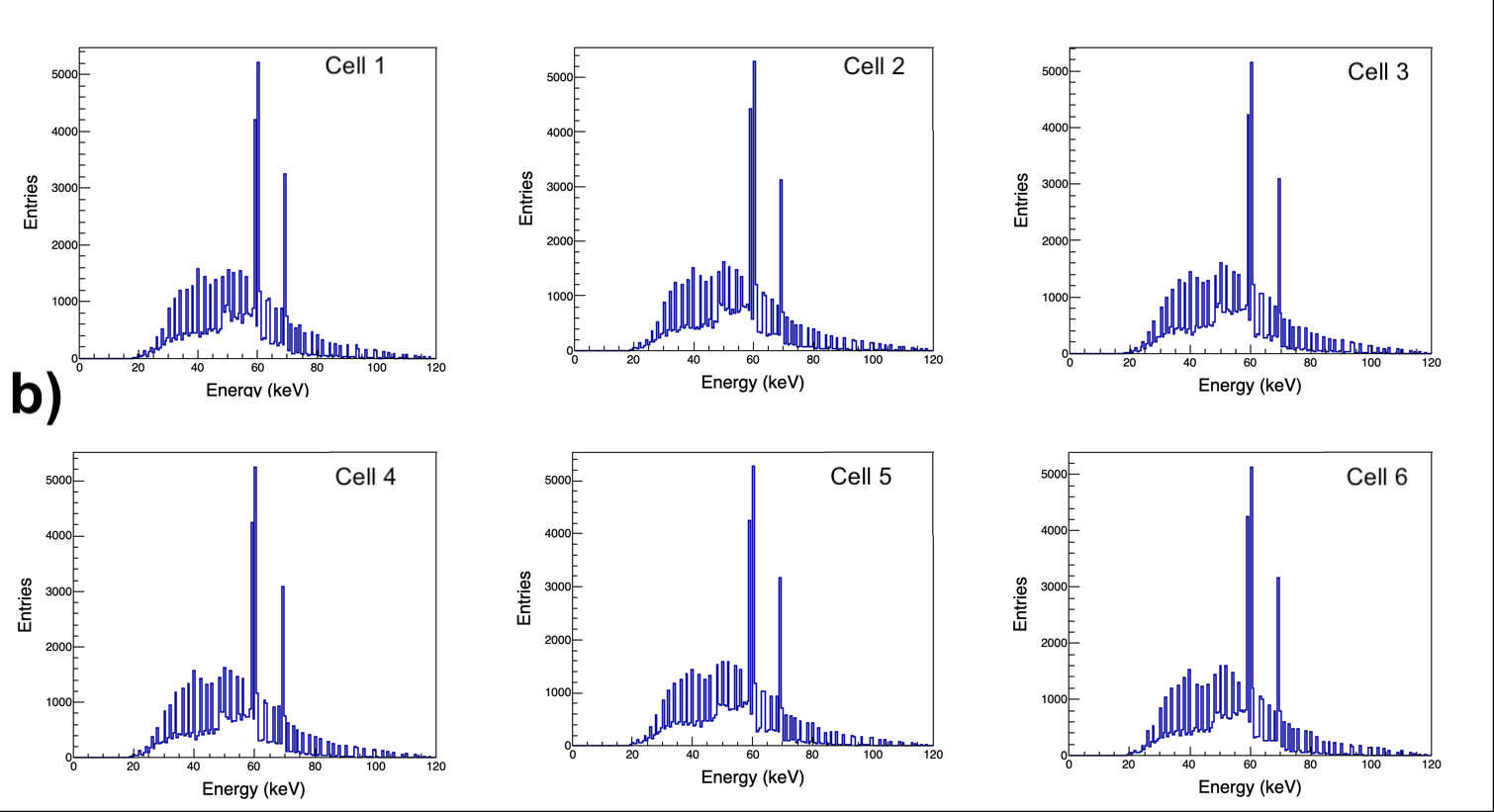}
        \caption{Energy distribution of X-rays detected in each hexagonal cell of the BMD for source-to-detector distances of a) 50 cm  and b) 200 cm (bottom).}
        \label{distribution}
      \end{figure}

\begin{figure}
        \includegraphics[width=\linewidth]{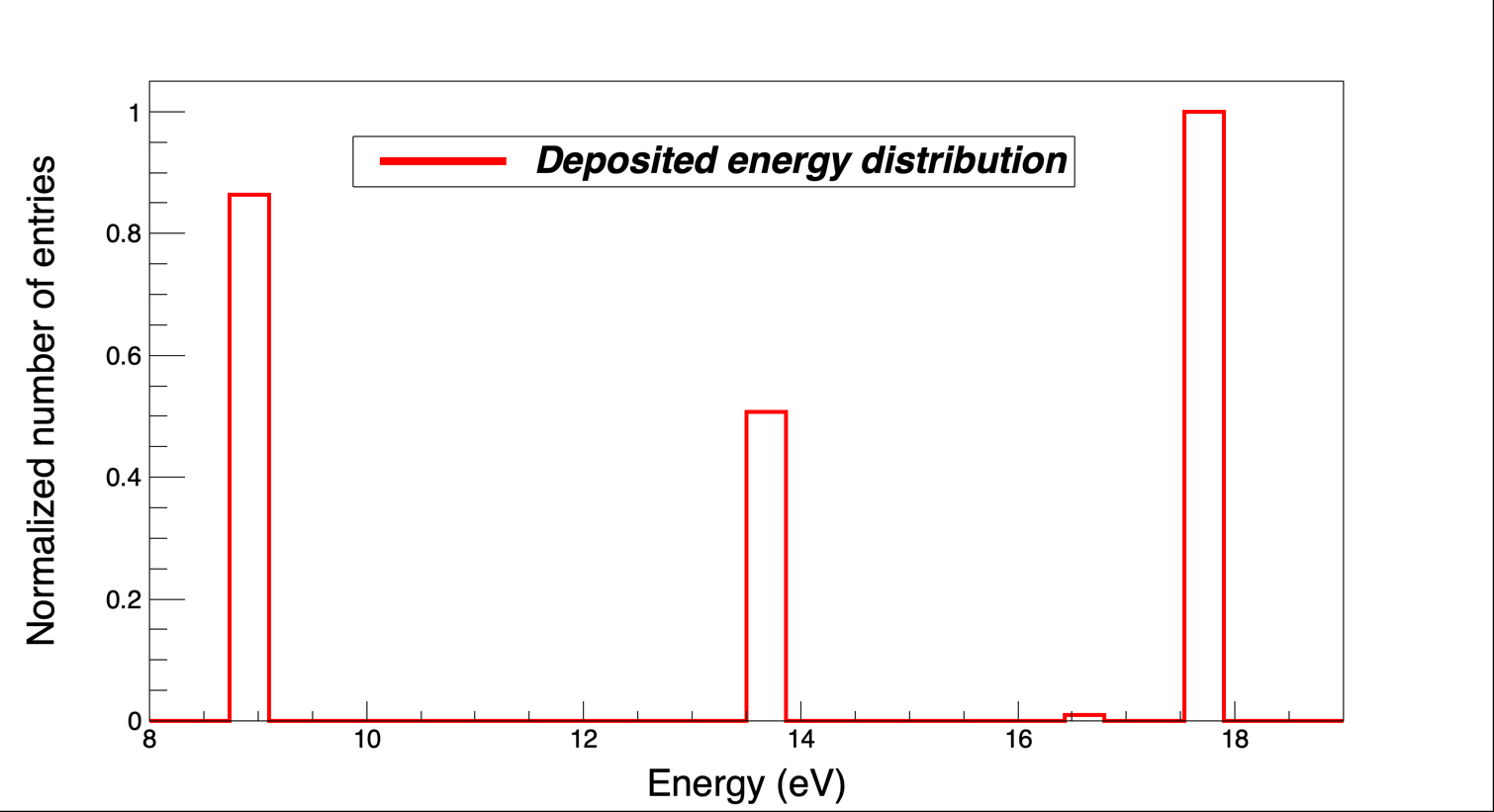}
        \caption{Distribution of the total energy deposited in the BMD arising from the tungsten X-ray spectrum.}
        \label{deposited}
      \end{figure}

  The X-ray energy was also evaluated in each cell and corresponds to the tungsten X-ray spectrum described previously. Figure~\ref{distribution} presents the energy distributions obtained at 50~cm and 200~cm. For a given distance, a homogeneous distribution of entries is observed among all cells. Figure~\ref{deposited} shows the total energy deposited in the BMD, and the same behavior is observed in all cells and at all simulated distances. Finally, the energy deposited in the detector support structure was calculated and found to be negligible compared with that deposited in the active detector volume.

  \section{Backscatter analysis}~\label{backscatter}
\begin{figure}
      \includegraphics[width=\linewidth]{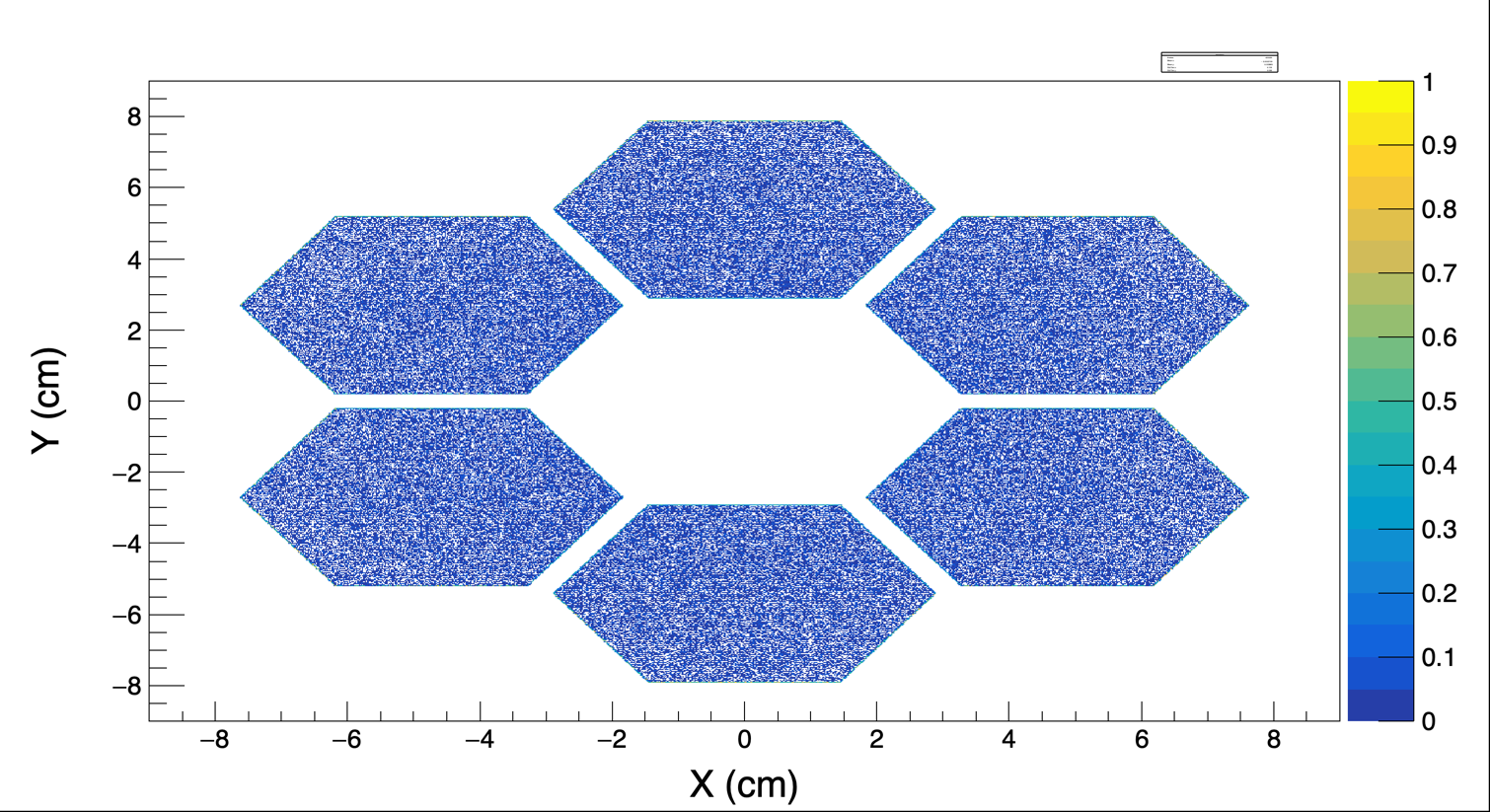}
      \caption{Backscattered X-ray hit multiplicity recorded in the BMD.}
      \label{backscatterhits}
    \end{figure}

  Understanding the behavior of backscattered X-rays is important for radiation protection. In this final analysis, a water cube with dimensions of $50 \times 50 \times 50$~cm$^3$ was placed 50~cm from the BMD, with the X-ray source located at the center of the detector so that the emitted X-rays passed through the BMD before reaching the water phantom. This configuration emulates the operation of a portable X-ray unit, in which the operator is typically positioned behind the device. The purpose of this study is to evaluate the contribution of backscattered radiation to the exposure of the operator.
Figure~\ref{backscatterhits} shows the multiplicity recorded in the BMD due to backscattered photons. The number of detected X-rays was compared with the multiplicities obtained at source-to-detector distances of 50, 100, 150, and 200~cm. The relative contributions are summarized in Table~\ref{percentage}. For the direct beam, the multiplicity decreases with distance according to the geometric spreading of the X-ray field. In contrast, the number of backscattered photons depends primarily on the size of the irradiated object and the probability of photon interactions within it. Because the water phantom has dimensions of $50 \times 50$~cm$^2$ in the transverse plane, it intercepts a large fraction of the incident photons and scatters a significant number of them toward the detector. As a result, the number of photons detected in the backward direction exceeds that observed for the direct beam at larger source-to-detector distances.

    \begin{table}
      \centering
      \caption{Percentage of X-rays detected by the BMD at five different position}
      \begin{tabular}{c | c }
        BMD location & percentage detected\\
        \hline
        50~cm&  2.37\%\\
        \hline
        100~cm&  0.61\%\\
        \hline
        150~cm&  0.27\%\\
        \hline
        200~cm&  0.15\%\\
        \hline
        backscatter&  6.49\%\\
      \end{tabular}\label{percentage}
    \end{table}

\begin{figure}
      \includegraphics[width=\linewidth]{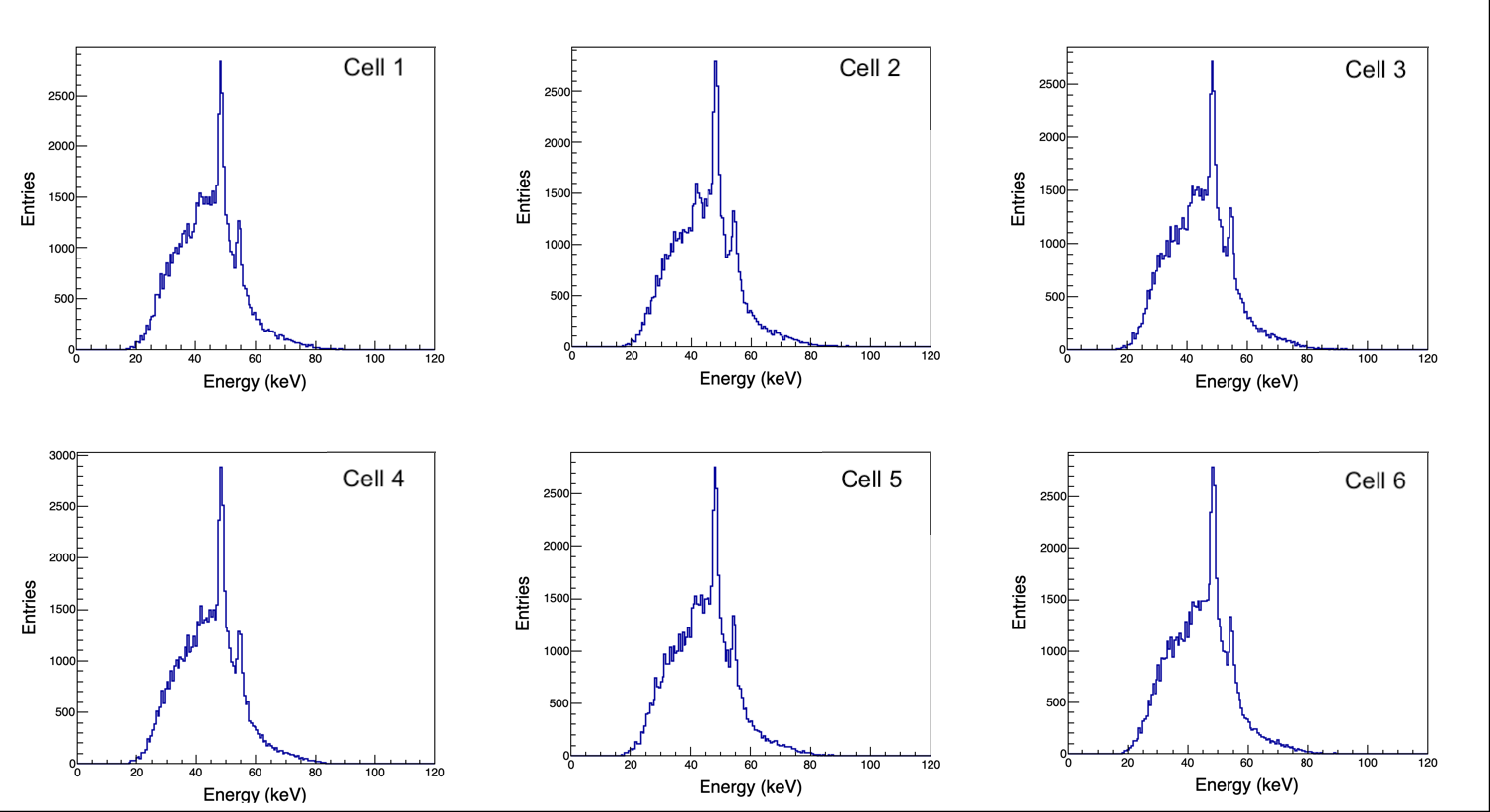}
      \caption{Energy distribution of backscattered X-rays detected in each hexagonal cell of the BMD.}
      \label{scatterenergy}
    \end{figure}

The X-ray energy was measured in each cell, and the resulting energy distribution is shown in Figure~\ref{scatterenergy}. Photons with energies above 100~keV were found to contribute negligibly to the backscattered component, indicating that high-energy photons are predominantly transmitted through the water phantom without significant scattering toward the detector. Finally, the total energy deposited in the BMD exhibits the same behavior observed in Figure~\ref{deposited}, confirming that the deposited energy distribution remains consistent across the different measurement conditions and detector positions.\\
A controlled study using a water phantom and the proposed beam monitoring detector would allow the simulated backscatter fraction and spatial distribution to be compared with measured data. Such a comparison would help quantify the accuracy of the simulation model, assess the influence of additional experimental effects, and confirm the applicability of the proposed detector for scatter radiation studies and radiological protection.

\section{Uncertainty and Limitations}\label{uncertainty}
The uncertainties reported throughout this work correspond to the statistical uncertainties obtained from the fitting procedures used to extract the relevant detector parameters. In addition to these statistical contributions, the simulated detector response is subject to systematic uncertainties associated with the assumptions adopted in the Monte Carlo model. These include the simplified X-ray source description, idealized spatial and angular beam distributions, uncertainties in the optical properties of the scintillator and reflector materials, and the exclusion of photodetector and electronic effects such as photon detection efficiency, transit time spread, and electronic jitter.
Particular attention should be given to the optical properties of the reflector materials, since the ITR was found to depend on the scintillator wrapping. Variations in reflector type, surface finish, reflectivity, and optical coupling conditions may modify the photon transport and, consequently, the absolute ITR values predicted by the simulation.\\
The identification of these systematic uncertainties provides a framework for estimating the range over which the predicted detector performance may vary under more realistic operating conditions. In particular, variations in source geometry, beam divergence, optical parameters, and photosensor characteristics may affect the absolute values of quantities such as detection efficiency and ITR. Quantifying the influence of these parameters in future studies will make it possible to establish confidence intervals for the expected detector response and to identify the parameters to which the system is most sensitive, thereby guiding further optimization of the proposed X-ray beam monitor.
Although these factors may modify the absolute numerical values reported here, the overall trends and the relative response of the six scintillator cells are expected to remain robust. Consequently, the present results should be interpreted as a first-order estimate of the performance of the proposed detector concept.

\section{Conclusions}\label{conclusions}

This work presents a Monte Carlo simulation study of a portable X-ray beam monitoring detector (BMD) designed to provide real-time information on beam intensity, spatial uniformity, and backscattered radiation, while supporting radiation protection in portable X-ray applications. The proposed detector consists of six hexagonal BC404 plastic scintillators, each with a height of 5~cm and a thickness of 2~cm.\\
For a tungsten X-ray spectrum in the energy range from 20 to 120~keV, the simulated detector exhibited an ITR of 100.8 $\pm$ 2.1~ps.  The optical analysis showed that the reflector reflectivity has a significant influence on the ITR of the detector. Although the ideal 100\% reflective wrapping provided the highest light collection efficiency, the 93\% reflective material yielded a slightly lower ITR than the ideal case. 
Under the simulation conditions considered in this work, the use of a reflectivity of 93\% yielded a lower ITR than the idealized 100\% case. This behavior is attributed to the preferential loss of late-arriving photons at each reflection, which reduces the temporal spread of the arrival-time distribution. Additional simulation and experimental studies are required to confirm this trend and to determine the optimal reflector configuration for this application. 
The results also showed that the X-ray multiplicity decreases with increasing distance from the source, following the expected inverse-square law. The energy distributions were homogeneous among the six cells, and the deposited energy exhibited the same general behavior at all simulated distances.\\
The simulations further showed that photons with energies above 100~keV contribute negligibly to the backscattered component, indicating that high-energy photons are predominantly transmitted through the scattering medium. A backscatter analysis using a water phantom demonstrated that approximately 6\% of the incident beam was detected as backscattered radiation. In addition, the number of backscattered photons exceeded the number of directly detected photons at large source-to-detector distances, emphasizing the importance of scattered radiation in radiation protection studies.\\
Plastic scintillators provide a cost-effective, mechanically robust, and fast-response alternative to inorganic crystal scintillators. The compact dimensions of the proposed BMD make it suitable for use in confined spaces where radiation measurements are required, while the hexagonal geometry provides high granularity and allows the active area to be expanded by adding additional cells.\\
An effective beam monitor should perturb the incident beam as little as possible. According to the present simulations, the proposed BMD interacts with approximately 10\% of the incident photons, indicating that it can operate as a practical beam monitor while preserving most of the beam intensity. These results will be validated in future experimental studies.\\
The results presented in this work are based entirely on Monte Carlo simulations and should therefore be regarded as a first-order estimate of the expected detector performance. Experimental measurements are required to validate the predicted intrinsic time resolution, detection efficiency, and backscatter response under realistic operating conditions. Future work will focus on extending the analysis to different scattering materials, alternative beam geometries, and specific operating-room configurations. These studies will provide a more comprehensive characterization of the radiation field and further assess the versatility of the proposed detector design.

\printbibliography

@article{Korpachev_2017,
doi = {10.1088/1742-6596/798/1/012218},
url = {https://doi.org/10.1088/1742-6596/798/1/012218},
year = {2017},
month = {jan},
publisher = {IOP Publishing},
volume = {798},
number = {1},
pages = {012218},
author = {Korpachev, S and Chadeeva, M},
title = {Geant4 simulation of optical photon transport in scintillator tile with direct readout by silicon photomultiplier},
journal = {Journal of Physics: Conference Series}
}

@techreport{BC404,
  author = {Saint-Gobain C.},
  title = {BC400 BC404 BC408 BC412 BC416 Data
Sheet },
  year = {2018}
}

@techreport{lightouput,
  author = {Wieczorek A.},
  title = {Development of novel plastic scintillators based
on polyvinyltoluene for the hybrid J-PET/MR tomograph},
  year = {2017}
}

@article{tungsten,
title = {X-ray spectra and gamma factors from 70 to 120 kV X-ray tube voltages},
journal = {Radiation Physics and Chemistry},
volume = {184},
pages = {109437},
year = {2021},
issn = {0969-806X},
doi = {https://doi.org/10.1016/j.radphyschem.2021.109437},
url = {https://www.sciencedirect.com/science/article/pii/S0969806X21000876},
author = {Guillermo Eduardo Campillo-Rivera and Carina Oliva Torres-Cortes and Joel Vazquez-Bañuelos and Mayra Guadalupe Garcia-Reyna and Claudia Angelica Marquez-Mata and Marcial Vasquez-Arteaga and Hector Rene Vega-Carrillo}
}

@article{Jacquet2016,
  author       = {Jacquet, M.},
  title        = {Potential of compact Compton sources in the medical field},
  journal      = {Physica Medica},
  year         = {2016},
  volume       = {32},
  number       = {9},
  pages        = {1285--1294},
  doi          = {10.1016/j.ejmp.2016.09.011}
}

@article{Eggl2017,
  author       = {Eggl, E. and Schleede, S. and Bech, M. and Achterhold, K. and Grandl, S. and Sztrókay, A. and Hellerhoff, K. and Mayr, D. and Loewen, R. and Ruth, R.~D. and Reiser, M.~F. and Pfeiffer, F.},
  title        = {X-ray phase-contrast tomosynthesis of a human ex vivo breast slice with an inverse Compton x-ray source},
  journal      = {EPL},
  year         = {2017},
  volume       = {116},
  pages        = {1--7},
  articleNumber= {68003},
  doi          = {10.1209/0295-5075/116/68003}
}

@article{Paterno2020,
  author       = {Paternò, G. and Cardarelli, P. and Gambaccini, M. and Taibi, A.},
  title        = {Dual-Energy X-ray Medical Imaging with Inverse Compton Sources: A Simulation Study},
  journal      = {Crystals},
  year         = {2020},
  volume       = {10},
  number       = {9},
  pages        = {834},
  doi          = {10.3390/cryst10090834}
}

@article{Kulpe2020,
  author       = {Kulpe, S. and others},
  title        = {Spectroscopic imaging at compact inverse Compton X‐ray sources},
  journal      = {Physics in Medicine \& Biology},
  year         = {2020},
  volume       = {65},
  number       = {24},
  pages        = {24},
  doi          = {10.1088/1361-6560/abccxx}
}

@article{Huang2021,
  author       = {Huang, Juanjuan and Deng, Fuli and G{\"u}nther, Benedikt and Achterhold, Klaus and Liu, Yue and Jentys, Andreas and Lercher, Johannes A. and Dierolf, Martin and Pfeiffer, Franz},
  title        = {Laboratory-scale: In situ X-ray absorption spectroscopy of a palladium catalyst on a compact inverse-Compton scattering X-ray beamline},
  journal      = {Journal of Analytical Atomic Spectrometry},
  year         = {2021},
  volume       = {36},
  number       = {12},
  pages        = {2649--2659},
  doi          = {10.1039/d1ja00274k}
}

@article{Huang2020,
  author       = {Huang, Juanjuan and G{\"u}nther, Benedikt and Achterhold, Klaus and Cui, Yi tao and Gleich, Bernhard and Dierolf, Martin and Pfeiffer, Franz},
  title        = {Energy-Dispersive X-ray Absorption Spectroscopy with an Inverse Compton Source},
  journal      = {Scientific Reports},
  year         = {2020},
  volume       = {10},
  articleNumber= {8772},
  doi          = {10.1038/s41598-020-65225-4}
}

@article{Melcher2024,
  author       = {Melcher, Johannes and Dierolf, Martin and G{\"u}nther, Benedikt and Achterhold, Klaus and Pfeiffer, Franz},
  title        = {High-energy X-ray diffraction experiment employing a compact synchrotron X-ray source based on inverse Compton scattering},
  journal      = {Zeitschrift f{\"u}r Medizinische Physik},
  year         = {2024},
  doi          = {10.1016/j.zemedi.2024.03.003}
}

@article{Franck2021,
  author       = {Franck, Carl and Muratori, Bruno D. and Williams, Peter H. and Krafft, Geoffrey A. and Terzi{\'c}, Bal{\v s}a},
  title        = {Intense monochromatic photons above 100 keV from an inverse Compton source},
  journal      = {Physical Review Accelerators and Beams},
  year         = {2021},
  volume       = {24},
  articleNumber= {050701},
  doi          = {10.1103/PhysRevAccelBeams.24.050701}
}

@article{Habib2015,
  author       = {Habib, T. and others},
  title        = {Development and applications of Compton camera—a review},
  journal      = {Sensors (Basel)},
  year         = {2022},
  volume       = {22},
  number       = {21},
  pages        = {8539},
  doi          = {10.3390/s22218539}
}

@article{Hsieh2016,
  author       = {Hsieh, J. and others},
  title        = {Energy-dispersive CdTe and CdZnTe detectors for spectral clinical CT and NDT applications},
  journal      = {Physics in Medicine \& Biology},
  year         = {2015},
  volume       = {60},
  number       = {13},
  pages        = {5471--5486},
  doi          = {10.1088/0031-9155/60/13/5471}
}

@article{Crespo2016,
  author       = {Crespo, P. and others},
  title        = {CdTe based energy resolving, X-ray photon counting detector performance assessment: The effects of charge sharing correction algorithm choice},
  journal      = {Sensors},
  year         = {2020},
  volume       = {20},
  number       = {21},
  pages        = {6093},
  doi          = {10.3390/s20216093}
}

@article{Iniewski2011,
  author       = {Iniewski, K.},
  title        = {Photon counting energy dispersive detector arrays for X-ray imaging},
  journal      = {Nuclear Instruments and Methods in Physics Research A},
  year         = {2009},
  volume       = {607},
  number       = {1},
  pages        = {85--88},
  doi          = {10.1016/j.nima.2009.01.048}
}

@incollection{Gunther2023,
  author       = {Benedikt Sebastian Günther},
  title        = {X-ray Beam Position Monitoring and Stabilisation},
  booktitle    = {Storage Ring-Based Inverse Compton X-ray Sources},
  editor       = {Benedikt Sebastian Günther},
  series       = {Springer Theses},
  pages        = {243--253},
  year         = {2023},
  publisher    = {Springer},
  doi          = {10.1007/978-3-031-17742-2_10},
  url          = {https://link.springer.com/chapter/10.1007/978-3-031-17742-2_10}
}

@article{Gunther2019,
  author       = {Benedikt Günther and Martin Dierolf and Klaus Achterhold and Franz Pfeiffer},
  title        = {Device for source position stabilization and beam parameter monitoring at inverse Compton X-ray sources},
  journal      = {Journal of Synchrotron Radiation},
  year         = {2019},
  volume       = {26},
  number       = {5},
  pages        = {1546--1553},
  doi          = {10.1107/S1600577519006453},
  pmcid        = {PMC6730616},
  url          = {https://pmc.ncbi.nlm.nih.gov/articles/PMC6730616/}
}

@article{ALLEN2003549,
title = {PHENIX inner detectors},
journal = {Nuclear Instruments and Methods in Physics Research Section A: Accelerators, Spectrometers, Detectors and Associated Equipment},
volume = {499},
number = {2},
pages = {549-559},
year = {2003},
note = {The Relativistic Heavy Ion Collider Project: RHIC and its Detectors},
issn = {0168-9002},
doi = {https://doi.org/10.1016/S0168-9002(02)01956-3},
url = {https://www.sciencedirect.com/science/article/pii/S0168900202019563},
author = {M. Allen and M.J. Bennett and M. Bobrek and J.B. Boissevain and S. Boose and E. Bosze and C. Britton and J. Chang and C.Y. Chi and M. Chiu and R. Conway and R. Cunningham and A. Denisov and A. Deshpande and M.S. Emery and A. Enokizono and N. Ericson and B. Fox and S.-Y. Fung and P. Giannotti and T. Hachiya and A.G. Hansen and K. Homma and B.V. Jacak and D. Jaffe and J.H. Kang and J. Kapustinsky and S.Y. Kim and Y.G. Kim and T. Kohama and P.J. Kroon and W. Lenz and N. Longbotham and M. Musrock and T. Nakamura and H. Ohnishi and S.S. Ryu and A. Sakaguchi and R. Seto and T. Shiina and M. Simpson and J. Simon-Gillo and W.E. Sondheim and T. Sugitate and J.P. Sullivan and H.W. {van Hecke} and J.W. Walker and S.N. White and P. Willis and N. Xu}
}

@article{ALICECollaboration_2008,
doi = {10.1088/1748-0221/3/08/S08002},
url = {https://doi.org/10.1088/1748-0221/3/08/S08002},
year = {2008},
month = {aug},
publisher = {},
volume = {3},
number = {08},
pages = {S08002},
author = {The ALICE Collaboration and K Aamodt and A Abrahantes Quintana and R Achenbach and S Acounis and D Adamová and C Adler and M Aggarwal and F Agnese and G Aglieri Rinella and Z Ahammed and A Ahmad and N Ahmad and S Ahmad and A Akindinov and P Akishin and D Aleksandrov and B Alessandro and R Alfaro and G Alfarone and A Alici and J Alme and T Alt and S Altinpinar and W Amend and C Andrei and Y Andres and A Andronic and G Anelli and M Anfreville and V Angelov and A Anzo and C Anson and T Anticić and V Antonenko and D Antonczyk and F Antinori and S Antinori and P Antonioli and L Aphecetche and H Appelshäuser and V Aprodu and M Arba and S Arcelli and A Argentieri and N Armesto and R Arnaldi and A Arefiev and I Arsene and A Asryan and A Augustinus and T C Awes and J Äysto and M Danish Azmi and S Bablock and A Badalà and S K Badyal and J Baechler and S Bagnasco and R Bailhache and R Bala and A Baldisseri and A Baldit and J Bán and R Barbera and P-L Barberis and J M Barbet and G Barnäfoldi and V Barret and J Bartke and D Bartos and M Basile and V Basmanov and N Bastid and G Batigne and B Batyunya and J Baudot and C Baumann and I Bearden and B Becker and J Belikov and R Bellwied and E Belmont-Moreno and A Belogianni and S Belyaev and A Benato and J L Beney and L Benhabib and F Benotto and S Beolé and I Berceanu and A Bercuci and E Berdermann and Y Berdnikov and C Bernard and R Berny and J D Berst and H Bertelsen and L Betev and A Bhasin and P Baskar and A Bhati and N Bianchi and J Bielčik and J Bielčiková and L Bimbot and G Blanchard and F Blanco and F Blanco and D Blau and C Blume and S Blyth and M Boccioli and A Bogdanov and H Bøggild and M Bogolyubsky and L Boldizsár and M Bombara and C Bombonati and M Bondila and D Bonnet and V Bonvicini and H Borel and F Borotto and V Borshchov and Y Bortoli and O Borysov and S Bose and L Bosisio and M Botje and S Böttger and G Bourdaud and O Bourrion and S Bouvier and A Braem and M Braun and P Braun-Munzinger and L Bravina and M Bregant and G Bruckner and R Brun and E Bruna and O Brunasso and G E Bruno and D Bucher and V Budilov and D Budnikov and H Buesching and P Buncic and M Burns and S Burachas and O Busch and J Bushop and X Cai and H Caines and F Calaon and M Caldogno and I Cali and P Camerini and R Campagnolo and M Campbell and X Cao and G P Capitani and G Cara Romeo and M Cardenas-Montes and H Carduner and F Carena and W Carena and P Cariola and F Carminati and J Casado and A Casanova Diaz and M Caselle and J Castillo Castellanos and J Castor and V Catanescu and E Cattaruzza and D Cavazza and P Cerello and S Ceresa and V Černý and V Chambert and S Chapeland and A Charpy and D Charrier and M Chartoire and J L Charvet and S Chattopadhyay and S Chattopadhyay and V Chepurnov and S Chernenko and M Cherney and C Cheshkov and B Cheynis and P Chochula and E Chiavassa and V Chibante Barroso and J Choi and P Christakoglou and P Christiansen and C Christensen and O A Chykalov and C Cicalo and L Cifarelli-Strolin and M Ciobanu and F Cindolo and C Cirstoiu and O Clausse and J Cleymans and O Cobanoglu and J-P Coffin and S Coli and A Colla and C Colledani and C Combaret and M Combet and M Comets and G Conesa Balbastre and Z Conesa del Valle and G Contin and J Contreras and T Cormier and F Corsi and P Cortese and F Costa and E Crescio and P Crochet and E Cuautle and J Cussonneau and M Dahlinger and A Dainese and H H Dalsgaard and L Daniel and I Das and T Das and A Dash and R Da Silva and M Davenport and H Daues and A De Caro and G de Cataldo and J De Cuveland and A De Falco and M de Gaspari and P de Girolamo and J de Groot and D De Gruttola and A De Haas and N De Marco and S De Pasquale and P De Remigis and D de Vaux and G Decock and H Delagrange and M Del Franco and G Dellacasa and C Dell'Olio and D Dell'Olio and A Deloff and V Demanov and E Dénes and G D'Erasmo and D Derkach and A Devaux and D Di Bari and A Di Bartolomeo and C Di Giglio and S Di Liberto and A Di Mauro and P Di Nezza and M Dialinas and L Diaz and R Díaz Valdes and T Dietel and R Dima and H Ding and C Dinca and R Divià and V Dobretsov and A Dobrin and B Doenigus and T Dobrowolski and I Domínguez and M Dorn and S Drouet and A E Dubey and L Ducroux and F Dumitrache and E Dumonteil and P Dupieux and V Duta and A Dutta Majumdar and M Dutta Majumdar and Th Dyhre and L Efimov and A Efremov and D Elia and D Emschermann and C Engster and A Enokizono and B Espagnon and M Estienne and A Evangelista and D Evans and S Evrard and C W Fabjan and D Fabris and J Faivre and D Falchieri and A Fantoni and R Farano and R Fearick and O Fedorov and V Fekete and D Felea and G Feofilov and A Férnandez Téllez and A Ferretti and F Fichera and S Filchagin and E Filoni and C Finck and R Fini and E M Fiore and D Flierl and M Floris and Z Fodor and Y Foka and S Fokin and P Force and F Formenti and E Fragiacomo and M Fragkiadakis and D Fraissard and A Franco and M Franco and U Frankenfeld and U Fratino and S Fresneau and A Frolov and U Fuchs and J Fujita and C Furget and M Furini and M Fusco Girard and J-J Gaardhøje and A Gabrielli and S Gadrat and M Gagliardi and A Gago and L Gaido and A Gallas Torreira and M Gallio and E Gandolfi and P Ganoti and M Ganti and J Garabatos and A Garcia Lopez and L Garizzo and L Gaudichet and R Gemme and M Germain and A Gheata and M Gheata and B Ghidini and P Ghosh and G Giolu and G Giraudo and P Giubellino and R Glasow and P Glässel and E G Ferreiro and C Gonzalez Gutierrez and L H Gonzales-Trueba and S Gorbunov and Y Gorbunov and H Gos and J Gosset and S Gotovac and H Gottschlag and D Gottschalk and V Grabski and T Grassi and H Gray and O Grebenyuk and K Grebieszkow and C Gregory and C Grigoras and N Grion and V Grigoriev and A Grigoryan and C Grigoryan and S Grigoryan and Y Grishuk and P Gros and J Grosse-Oetringhaus and J-Y Grossiord and R Grosso and B Grynyov and C Guarnaccia and F Guber and F Guerin and R Guernane and M Guerzoni and A Guichard and M Guida and G Guilloux and H Gulkanyan and K Gulbrandsen and T Gunji and A Gupta and V Gupta and H-A Gustafsson and H Gutbrod and C Hadjidakis and M Haiduc and G Hamar and H Hamagaki and J Hamblen and J C Hansen and P Hardy and D Hatzifotiadou and J W Harris and M Hartig and A Harutyunyan and A Hayrapetyan and D Hasch and D Hasegan and J Hehner and N Heine and M Heinz and H Helstrup and A Herghelegiu and S Herlant and G Herrera Corral and N Herrmann and K Hetland and P Hille and H Hinke and B Hippolyte and M Hoch and H Hoebbel and H Hoedlmoser and T Horaguchi and M Horner and P Hristov and I Hřivnáčová and S Hu and C Hu Guo and T Humanic and A Hurtado and D S Hwang and J C Ianigro and M Idzik and S Igolkin and R Ilkaev and I Ilkiv and M Imhoff and P G Innocenti and E Ionescu and M Ippolitov and M Irfan and C Insa and M Inuzuka and C Ivan and A Ivanov and M Ivanov and V Ivanov and P Jacobs and A Jacholkowski and L Jančurová and R Janik and M Jasper and C Jena and L Jirden and D P Johnson and G T Jones and C Jorgensen and F Jouve and P Jovanović and A Junique and A Jusko and H Jung and W Jung and K Kadija and A Kamal and R Kamermans and S Kapusta and A Kaidalov and V Kakoyan and S Kalcher and E Kang and J Kapitan and V Kaplin and K Karadzhev and O Karavichev and T Karavicheva and E Karpechev and K Karpio and A Kazantsev and U Kebschull and R Keidel and M Mohsin Khan and A Khanzadeev and Y Kharlov and D Kikola and B Kileng and D Kim and D S Kim and D W Kim and H N Kim and J S Kim and S Kim and J B Kinson and S K Kiprich and I Kisel and S Kiselev and A Kisiel and T Kiss and V Kiworra and J Klay and C Klein Bösing and M Kliemant and A Klimov and A Klovning and A Kluge and R Kluit and S Kniege and R Kolevatov and T Kollegger and A Kolojvari and V Kondratiev and E Kornas and E Koshurnikov and I Kotov and R Kour and M Kowalski and S Kox and K Kozlov and I Králik and F Kramer and I Kraus and A Kravčáková and T Krawutschke and M Krivda and E Kryshen and Y Kucheriaev and A Kugler and C Kuhn and P Kuijer and L Kumar and N Kumar and P Kumpumaeki and A Kurepin and A N Kurepin and S Kushpil and V Kushpil and M Kutovsky and H Kvaerno and M Kweon and J-C Labbé and F Lackner and P Ladron de Guevara and V Lafage and P La Rocca and M Lamont and C Lara and D T Larsen and G Laurenti and C Lazzeroni and Y Le Bornec and N Le Bris and C Le Gailliard and V Lebedev and J Lecoq and K S Lee and S C Lee and F Lefévre and I Legrand and T Lehmann and L Leistam and P Lenoir and V Lenti and H Leon and I Leon Monzon and P Lévai and Q Li and X Li and F Librizzi and R Lietava and N Lindegaard and V Lindenstruth and C Lippmann and M Lisa and O M Listratenko and F Littel and Y Liu and J Lo and V Lobanov and V Loginov and M López Noriega and R López-Ramírez and E López Torres and P M Lorenzo and G Løvhøiden and S Lu and W Ludolphs and M Lunardon and L Luquin and S Lusso and J-R Lutz and M Luvisetto and V Lyapin and A Maevskaya and C Magureanu and A Mahajan and S Majahan and T Mahmoud and A Mairani and D Mahapatra and A Makarov and I Makhlyueva and M Malek and T Malkiewicz and D Mal'Kevich and P Malzacher and A Mamonov and C Manea and L K Mangotra and D Maniero and V Manko and F Manso and V Manzari and Y Mao and A Marcel and S Marchini and J Mareš and G V Margagliotti and A Margotti and A Marin and J-C Marin and D Marras and P Martinengo and M I Martínez and A Martinez-Davalos and G Martínez Garcia and S Martini and A Marzari Chiesa and C Marzocca and S Masciocchi and M Masera and M Masetti and N I Maslov and A Masoni and F Massera and M Mast and A Mastroserio and Z L Matthews and B Mayer and G Mazza and M D Mazzaro and A Mazzoni and F Meddi and E Meleshko and A Menchaca-Rocha and S Meneghini and M Meoni and J Mercado Perez and P Mereu and O Meunier and Y Miake and A Michalon and R Michinelli and N Miftakhov and M Mignone and K Mikhailov and J Milosevic and Y Minaev and F Minafra and A Mischke and D Miśkowiec and V Mitsyn and C Mitu and B Mohanty and D Moisa and L Molnar and M Mondal and N Mondal and L Montaño Zetina and M Monteno and M Morando and M Morel and S Moretto and Th Morhardt and A Morsch and T Moukhanova and M Mucchi and V Muccifora and E Mudnic and H Müller and W Müller and J Munoz and D Mura and L Musa and J F Muraz and A Musso and R Nania and B Nandi and E Nappi and F Navach and S Navin and T Nayak and S Nazarenko and G Nazarov and L Nellen and F Nendaz and A Nianine and M Nicassio and B S Nielsen and S Nikolaev and V Nikolic and S Nikulin and V Nikulin and B Nilsen and M Nitti and F Noferini and P Nomokonov and G Nooren and F Noto and D Nouais and A Nyiri and J Nystrand and G Odyniec and H Oeschler and M Oinonen and M Oldenburg and I Oleks and E K Olsen and V Onuchin and C Oppedisano and F Orsini and A Ortiz-Velázquez and C Oskamp and A Oskarsson and F Osmic and L Österman and I Otterlund and G Ovrebekk and K Oyama and M Pachr and P Pagano and G Paić and C Pajares and S Pal and S Pal and G Pálla and A Palmeri and G Pancaldi and R Panse and A Pantaleo and G S Pappalardo and B Pastirčák and C Pastore and O Patarakin and V Paticchio and G Patimo and A Pavlinov and T Pawlak and T Peitzmann and Y Pénichot and A Pepato and H Pereira and D Peresunko and C Perez and J Perez Griffo and D Perini and D Perrino and W Peryt and A Pesci and V Peskov and Y Pestov and A J Peters and V Petráček and A Petridis and M Petris and V Petrov and V Petrov and M Petrovici and J Peyré and S Piano and A Piccotti and P Pichot and C Piemonte and M Pikna and R Pilastrini and P Pillot and O Pinazza and B Pini and L Pinsky and V Pinto Morais and V Pismennaya and F Piuz and R Platt and M Ploskon and S Plumeri and J Pluta and T Pocheptsov and P Podesta and F Poggio and M Poghosyan and T Poghosyan and K Polák and B Polichtchouk and P Polozov and V Polyakov and B Pommeresch and F Pompei and A Pop and S Popescu and F Posa and V Pospíšil and B Potukuchi and J Pouthas and S Prasad and R Preghenella and F Prino and L Prodan and G Prono and M A Protsenko and C A Pruneau and A Przybyla and I Pshenichnov and G Puddu and P Pujahari and A Pulvirenti and A Punin and V Punin and J Putschke and J Quartieri and E Quercigh and I Rachevskaya and A Rachevski and A Rademakers and S Radomski and A Radu and J Rak and L Ramello and R Raniwala and S Raniwala and O B Rasmussen and J Rasson and V Razin and K Read and J Real and K Redlich and C Reichling and C Renard and G Renault and R Renfordt and A R Reolon and A Reshetin and J-P Revol and K Reygers and H Ricaud and L Riccati and R A Ricci and M Richter and P Riedler and L M Rigalleau and F Riggi and W Riegler and E Rindel and J Riso and A Rivetti and M Rizzi and V Rizzi and M Rodriguez Cahuantzi and K Røed and D Röhrich and S Román-López and M Romanato and R Romita and F Ronchetti and P Rosinsky and P Rosnet and S Rossegger and A Rossi and V Rostchin and F Rotondo and F Roukoutakis and S Rousseau and C Roy and D Roy and P Roy and L Royer and G Rubin and A Rubio and R Rui and I Rusanov and G Russo and V Ruuskanen and E Ryabinkin and A Rybicki and S Sadovsky and K Šafařík and R Sahoo and J Saini and P Saiz and S Salur and S Sambyal and V Samsonov and L Šándor and A Sandoval and H Sann and J-C Santiard and R Santo and R Santoro and G Sargsyan and P Saturnini and E Scapparone and F Scarlassara and B Schackert and C Schiaua and R Schicker and T Schioler and J D Schippers and C Schmidt and H Schmidt and R Schneider and K Schossmaier and J Schukraft and Y Schutz and K Schwarz and K Schweda and E Schyns and G Scioli and E Scomparin and H Snow and S Sedykh and G Segato and S Sellitto and F Semeria and S Senyukov and H Seppänen and S Serci and L Serkin and S Serra and T Sesselmann and A Sevcenco and I Sgura and G Shabratova and R Shahoyan and E Sharkov and S Sharma and K Shigaki and K Shileev and P Shukla and A Shurygin and M Shurygina and Y Sibiriak and E Siddi and T Siemiarczuk and M H Sigward and A Silenzi and D Silvermyr and R Silvestri and E Simili and V Simion and R Simon and L Simonetti and R Singaraju and V Singhal and B Sinha and T Sinha and M Siska and B Sitár and M Sitta and B Skaali and P Skowronski and M Slodkowski and N Smirnov and L Smykov and R Snellings and W Snoeys and C Soegaard and J Soerensen and O Sokolov and A Soldatov and A Soloviev and H Soltveit and R Soltz and W Sommer and C Soos and F Soramel and S Sorensen and D Soyk and M Spyropoulou-Stassinaki and J Stachel and F Staley and I Stan and A Stavinskiy and J Steckert and G Stefanini and G Stefanek and T Steinbeck and H Stelzer and E Stenlund and D Stocco and M Stockmeier and G Stoicea and P Stolpovsky and P Strmeň and J S Stutzmann and G Su and T Sugitate and M Šumbera and C Suire and T Susa and K Sushil Kumar and D Swoboda and J Symons and I Szarka and A Szostak and M Szuba and P Szymanski and M Tadel and C Tagridis and L Tan and D Tapia Takaki and H Taureg and A Tauro and M Tavlet and G Tejeda Munoz and J Thäder and R Tieulent and P Timmer and T Tolyhy and N Topilskaya and C Torcato de Matos and H Torii and L Toscano and F Tosello and A Tournaire and T Traczyk and G Tröger and W Tromeur and D Truesdale and W Trzaska and G Tsiledakis and E Tsilis and A Tsvetkov and M Turcato and R Turrisi and M Tuveri and T Tveter and H Tydesjo and L Tykarski and K Tywoniuk and E Ugolini and K Ullaland and J Urbán and G M Urciuoli and G L Usai and M Usseglio and A Vacchi and M Vala and F Valiev and P Vande Vyvre and A Van Den Brink and N Van Eijndhoven and N Van Der Kolk and M van Leeuwen and L Vannucci and S Vanzetto and J-P Vanuxem and M A Vargas and R Varma and A Vascotto and A Vasiliev and M Vassiliou and P Vasta and V Vechernin and M Venaruzzo and E Vercellin and S Vergara and W Verhoeven and F Veronese and I Vetlitskiy and R Vernet and V Victorov and L Vidak and G Viesti and O Vikhlyantsev and Z Vilakazi and O Villalobos Baillie and A Vinogradov and L Vinogradov and Y Vinogradov and T Virgili and Y Viyogi and A Vodopianov and G Volpe and D Vranic and J Vrláková and B Vulpescu and C Wabnitz and V Wagner and L Wallet and R Wan and Y Wang and Y Wang and R Wheadon and R Weis and Q Wen and J Wessels and J Westergaard and J Wiechula and A Wiesenaecker and J Wikne and A Wilk and G Wilk and C Williams and N Willis and B Windelband and R Witt and H Woehri and K Wyllie and C Xu and C Yang and H Yang and F Yermia and Z Yin and Z Yin and B Yun Ky and I Yushmanov and B Yuting and E Zabrodin and S Zagato and B Zagreev and P Zaharia and A Zalite and G Zampa and C Zampolli and Y Zanevskiy and A Zarochentsev and O Zaudtke and P Závada and H Zbroszczyk and A Zepeda and V Zeter and I Zgura and M Zhalov and D Zhou and S Zhou and G Zhu and A Zichichi and A Zinchenko and G Zinovjev and Y Zoccarato and A Zubarev and A Zucchini and M Zuffa},
title = {The ALICE experiment at the CERN LHC},
journal = {Journal of Instrumentation}
}

@article{HARDING1999229,
title = {Coherent X-ray scatter imaging and its applications in biomedical science and industry},
journal = {Radiation Physics and Chemistry},
volume = {56},
number = {1},
pages = {229-245},
year = {1999},
issn = {0969-806X},
doi = {https://doi.org/10.1016/S0969-806X(99)00283-2},
url = {https://www.sciencedirect.com/science/article/pii/S0969806X99002832},
author = {G. Harding and B. Schreiber}
}

@article {Seibert3,
	author = {Seibert, J. Anthony and Boone, John M.},
	title = {X-Ray Imaging Physics for Nuclear Medicine Technologists. Part 2: X-Ray Interactions and Image Formation},
	volume = {33},
	number = {1},
	pages = {3--18},
	year = {2005},
	publisher = {Society of Nuclear Medicine},
	issn = {0091-4916},
	URL = {https://tech.snmjournals.org/content/33/1/3},
	eprint = {https://tech.snmjournals.org/content/33/1/3.full.pdf},
	journal = {Journal of Nuclear Medicine Technology}
}

@article{bebe,
title = {A beam–beam monitoring detector for the MPD experiment at NICA},
journal = {Nuclear Instruments and Methods in Physics Research Section A: Accelerators, Spectrometers, Detectors and Associated Equipment},
volume = {953},
pages = {163150},
year = {2020},
issn = {0168-9002},
doi = {https://doi.org/10.1016/j.nima.2019.163150},
url = {https://www.sciencedirect.com/science/article/pii/S0168900219314639},
author = {Mauricio Alvarado and Alejandro Ayala and Marco Alberto Ayala-Torres and Wolfgang Bietenholz and Isabel Dominguez and Marcos Fontaine and P. González-Zamora and Luis Manuel Montaño and E. Moreno-Barbosa and Miguel Enrique Patiño Salazar and L.A.P. Moreno and P.A. Nieto-Marín and V.Z. {Reyna Ortiz} and M. Rodríguez-Cahuantzi and G. Tejeda-Muñoz and Maria Elena Tejeda-Yeomans and A. Villatoro-Tello and C.H. {Zepeda Fernández}}
}

@techreport{Salvat2019PENELOPE2018,
  author       = {Nuclear Energy Agency (OECD)},
  title        = {PENELOPE–2018: A Code System for Monte Carlo Simulation of Electron and Photon Transport},
  institution  = {OECD Publishing},
  year         = {2019},
  doi          = {10.1787/32da5043-en},
  url          = {https://doi.org/10.1787/32da5043-en}
}

@article{Sloth2024McXtraceGPU,
  author       = {Sloth, Steffen and Willendrup, Peter Kj{\ae}r and Brandenborg S{\o}rensen, Hans Henrik and Christensen, Morten and Friis Poulsen, Henning},
  title        = {Accelerated ray‑tracing simulations using McXtrace},
  journal      = {arXiv preprint arXiv:2410.08747},
  year         = {2024},
  url          = {https://arxiv.org/abs/2410.08747}
}

@article{Kochebina2024OpenGATE,
  author       = {Kochebina, Olga and Bonifacio, Daniel A. B. and Konstantinou, Georgios and Paillet, Adrien and Pommranz, Christian M. and Razdevšek, Gašper and Sharyy, Viatcheslav and Yvon, Dominique and Jan, S{\'e}bastien},
  title        = {New GATE digitizer unit for versions post v9.3},
  journal      = {Frontiers in Physics},
  year         = {2024},
  volume       = {12},
  articleNumber= {1294916},
  doi          = {10.3389/fphy.2024.1294916},
  url          = {https://www.frontiersin.org/articles/10.3389/fphy.2024.1294916/full}
}

@phdthesis{Marquez2024,
  author       = {Márquez Quintos, E.},
  title        = {Simulación y caracterización de una celda para un monitor de radiación con fuentes de bajas energías},
  school       = {Benemérita Universidad Autónoma de Puebla},
  year         = {2024},
  type         = {Tesis doctoral},
  url          = {https://hdl.handle.net/20.500.12371/23740}
}

@article{LHCb,
  author       = {Ilgner, C. and Domke, M. and Lieng, M. and Nedos, M. and Sauerbrey, J. and Schleich, S. and Spaan, B. and Warda, K. and Wishahi, J.},
  title        = {The Beam Conditions Monitor of the LHCb Experiment},
  journal      = {IEEE Transactions on Nuclear Science},
  year         = {2010},
  volume       = {57},
  number       = {2},
  pages        = {814--822},
  doi          = {10.1109/TNS.2010.2047744},
  eprint       = {arXiv:1001.2487},
  url          = {https://arxiv.org/abs/1001.2487}
}

@inproceedings{ATLAS,
  author       = {Gorišek, A. and others},
  title        = {The ATLAS Beam Condition Monitor Commissioning},
  booktitle    = {Proceedings of the 13th International Conference on Radiation Effects on Semiconductor Devices, 2008},
  year         = {2008},
  note         = {CERN-CDS record no. 1158638, https://cds.cern.ch/record/1158638/files/p264.pdf},
  url          = {https://cds.cern.ch/record/1158638/files/p264.pdf}
}

@misc{SensL,
  author       = {onsemi (SensL C-Series)},
  title        = {MicroFC-60035-SMT (C-Series Blue-Sensitive SiPM) Datasheet},
  %howpublished =
  url={https://www.onsemi.com/pdf/datasheet/microc-series-d.pdf},
  year         = {2022}
  %note         = {Part number MICROFC-60035-SMT-TR1, Active area 6 mm×6 mm; 35 µm microcells; peak wavelength 420 nm.}
}

@article{Geant4,
  author       = {Agostinelli, S. and Allison, J. and Amako, K. and Apostolakis, J. and Araujo, H. and Arce, P. and Asai, M. and Axen, D. and Banerjee, S. and Barrand, G. and Behner, F. and Bellagamba, L. and Boudreau, J. and Broglia, L. and Brunengo, A. and Burkhardt, H. and Chauvie, S. and Chuma, J. and Chytracek, R. and Cooperman, G. and Cosmo, G. and Degtyarenko, P. and Dell’Acqua, A. and Depaola, G. and Dietrich, D. and Enami, R. and Feliciello, A. and Ferguson, C. and Fesefeldt, H. and Folger, G. and Foppiano, F. and Forti, A. and Garelli, S. and Giani, S. and Giannitrapani, R. and Gibin, D. and Gómez Cadenas, J. J. and González, I. and Gracia Abril, G. and Greiner, W. and Grichine, V. and Grossheim, A. and Guatelli, S. and Gumplinger, P. and Hamatsu, R. and Hashimoto, K. and Hasui, H. and Heikkinen, A. and Howard, A. and Ivanchenko, V. and Johnson, A. and Jones, F. W. and Kallenbach, J. and Kanaya, N. and Kawabata, M. and Kawabata, Y. and Kawaguti, M. and Kelner, S. and Kent, P. and Kimura, A. and Kodama, T. and Kokoulin, R. and Kossov, M. and Kurashige, H. and Lamanna, E. and Lampen, T. and Lara, V. and Lefebure, V. and Lei, F. and Liendl, M. and Lockman, W. and Longo, F. and Magni, S. and Maire, M. and Medernach, E. and Minamimoto, K. and Mora de Freitas, P. and Morita, Y. and Murakami, K. and Nagamatu, M. and Nartallo, R. and Nieminen, P. and Nishimura, T. and Ohtsubo, K. and Okamura, M. and O'Neale, S. and Oohata, Y. and Paech, K. and Perl, J. and Pfeiffer, A. and Pia, M. G. and Ranjard, F. and Rybin, A. and Sadilov, S. and Di Salvo, E. and Santin, G. and Sasaki, T. and Savvas, N. and Sawada, Y. and Scherer, S. and Sei, S. and Sirotenko, V. and Smith, D. and Starkov, N. and Stöcker, H. and Sulkimo, J. and Takahata, M. and Tanaka, S. and Tcherniaev, E. and Safai Tehrani, E. and Tropeano, M. and Truscott, P. and Uno, H. and Urban, L. and Urban, P. and Verderi, M. and Walkden, A. and Wander, W. and Weber, H. and Wellisch, J. P. and Wenaus, T. and Williams, D. C. and Wright, D. and Yamada, T. and Yoshida, H. and Zschiesche, D.},
  title        = {GEANT4 - a simulation toolkit},
  journal      = {Nuclear Instruments and Methods in Physics Research Section A},
  volume       = {506},
  number       = {3},
  pages        = {250--303},
  year         = {2003},
  doi          = {10.1016/S0168-9002(03)01368-8},
  url          = {https://doi.org/10.1016/S0168-9002(03)01368-8}
}

@article{Ranakoti2022_PLA_review,
AUTHOR = {Ranakoti, Lalit and Gangil, Brijesh and Mishra, Sandip Kumar and Singh, Tej and Sharma, Shubham and Ilyas, R.A. and El-Khatib, Samah},
TITLE = {Critical Review on Polylactic Acid: Properties, Structure, Processing, Biocomposites, and Nanocomposites},
JOURNAL = {Materials},
VOLUME = {15},
YEAR = {2022},
NUMBER = {12},
ARTICLE-NUMBER = {4312},
URL = {https://www.mdpi.com/1996-1944/15/12/4312},
PubMedID = {35744371},
ISSN = {1996-1944},
DOI = {10.3390/ma15124312}
}

@article{Chatzisavvas2022XrayTubeMC,
  author  = {Chatzisavvas, N. and Papadakis, E. and Politis, G.},
  title   = {Simulating Medical Imaging X-Ray Tubes with Various Anode Materials Using the Monte Carlo Method},
  journal = {Open Journal of Radiology},
  year    = {2022},
  volume  = {12},
  pages   = {171--186},
  doi     = {10.4236/ojrad.2022.123016}
}
\end{document}